\documentclass[twocolumn,aps,prb,superscriptaddress,longbibliography]{revtex4-2}
\usepackage{graphicx}
\usepackage{color}
\usepackage{amsmath}
\usepackage[export]{adjustbox}
\usepackage{enumitem}
\usepackage{amssymb}
\usepackage{hyperref}
\usepackage[normalem]{ulem}
\usepackage{adjustbox}

\hypersetup{
	colorlinks = true,
	linkcolor = blue,
	citecolor = blue,
	urlcolor  = blue,
}

\newcommand{\be}{\begin{equation}}
	\newcommand{\ee}{\end{equation}}

\newcommand{\bea}{\begin{eqnarray}}
	\newcommand{\eea}{\end{eqnarray}}

\newcommand{\p}{\partial}

\renewcommand{\vec}[1]{{\boldsymbol #1}}
\renewcommand{\epsilon}{\varepsilon}

\def\nn{\nonumber\\}

\begin{document}
	\title{Landau quantization near generalized Van Hove singularities: Magnetic breakdown and orbit networks}

	\author{V. A. Zakharov}
	\affiliation{Instituut-Lorentz, Universiteit Leiden, P.O. Box 9506, 2300 RA Leiden, The Netherlands}
	\author{A. Mert Bozkurt}
	\affiliation{Kavli Institute of Nanoscience, Delft University of Technology, Delft 2600 GA, The Netherlands}
	\affiliation{QuTech, Delft University of Technology, Delft 2600 GA, The Netherlands}
	\author{A. R. Akhmerov}
	\affiliation{Kavli Institute of Nanoscience, Delft University of Technology, Delft 2600 GA, The Netherlands}
	\author{D. O. Oriekhov}
	\affiliation{Instituut-Lorentz, Universiteit Leiden, P.O. Box 9506, 2300 RA Leiden, The Netherlands}
	
	\begin{abstract}
		We develop a theory of magnetic breakdown (MB) near high-order saddle points in the dispersions of two-dimensional materials, where two or more semiclassical cyclotron orbits approach each other. MB occurs due to quantum tunneling between several trajectories, which leads to non-trivial scattering amplitudes and phases. We show that for any saddle point this problem can be solved by mapping it to a scattering problem in a 1D tight-binding chain. Moreover, the occurrence of magnetic breakdown on the edges of the Brillouin zone facilitates the delocalization of the bulk Landau level states and the formation of 2D orbit networks. These extended network states compose dispersive mini-bands with finite energy broadening. This effect can be observed in transport experiments as a strong enhancement of the longitudinal bulk conductance in a quantum Hall bar. In addition, it may be probed in STM experiments by visualizing bulk current patterns.
	\end{abstract}
	\maketitle
	
	Magnetic breakdown (MB) in a single Bloch band occurs when two semiclassical trajectories of quasiparticles come close to each other and quantum tunneling between them becomes possible. This situation naturally appears near usual saddle points that give rise to logarithmic van Hove singularities in the density of states \cite{vanHove1953}. In novel atomically-thin 2D materials a new family of saddle points arises, around which the dispersion is flatter than in the usual case. This leads to power-law divergences in the density of states known as high-order van Hove singularities \cite{Yuan2019Nature,Chamon2020,Yuan2020PRB}. In some cases, more than two trajectories approach the saddle point, creating a MB structure with a larger s-matrix size proportional to the number of trajectories. In this paper, we present a method to calculate the precise MB s-matrix for any type of saddle point. It is based on rewriting the effective Hamiltonian in the Landau level basis, mapping the resulting algebraic problem to the 1D scattering in the quantum chain, and calculating the MB s-matrix by properly fixing semiclassical modes in the far-away region.
	
	\begin{figure}[h!]
		\centering
		\includegraphics[scale=0.38]{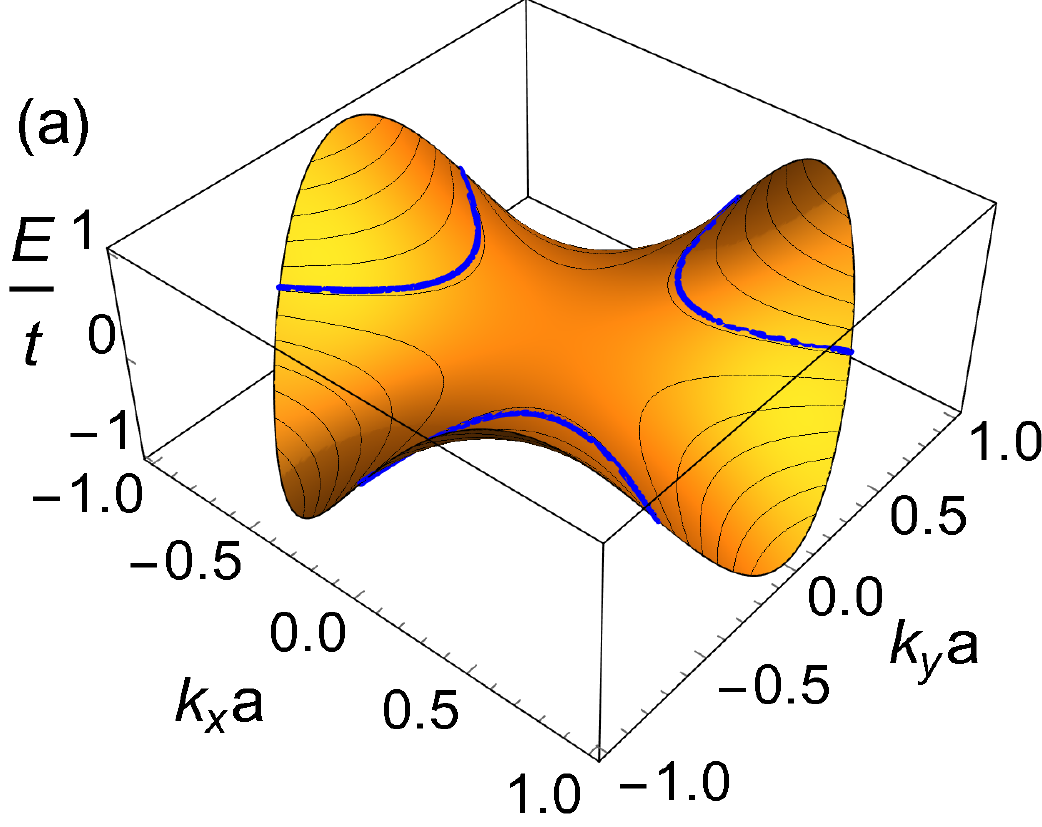}
		\includegraphics[scale=0.38]{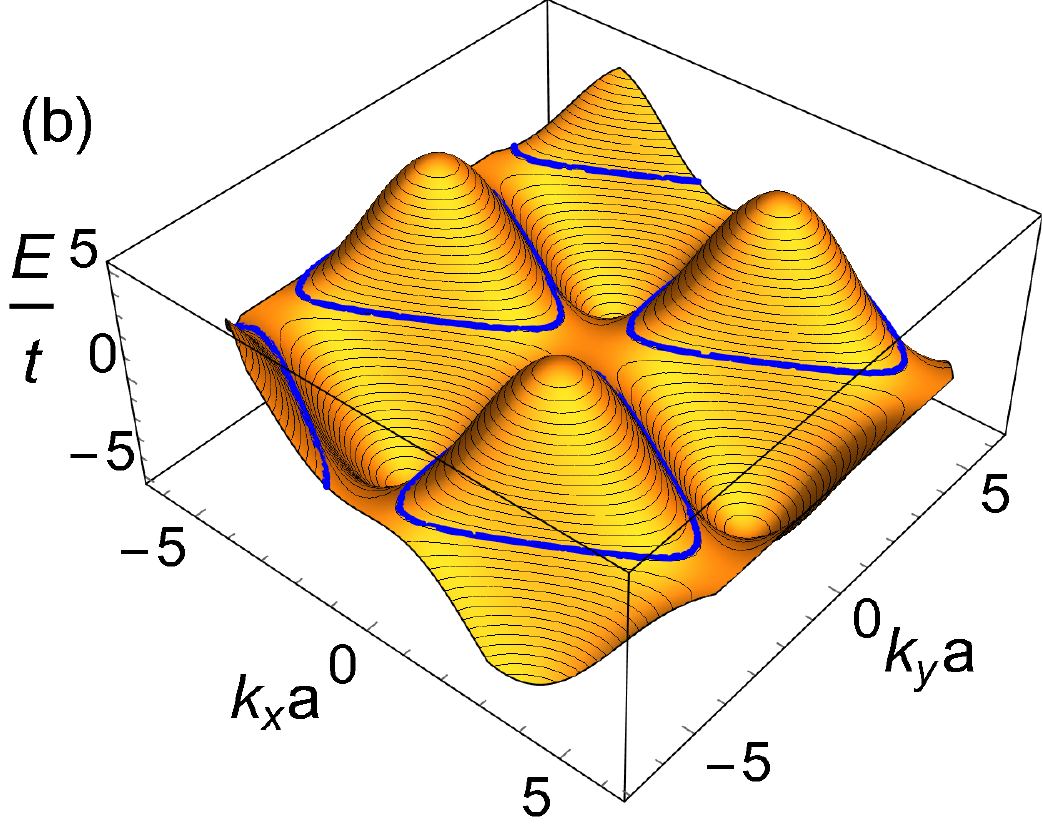}
		\includegraphics[scale=0.073]{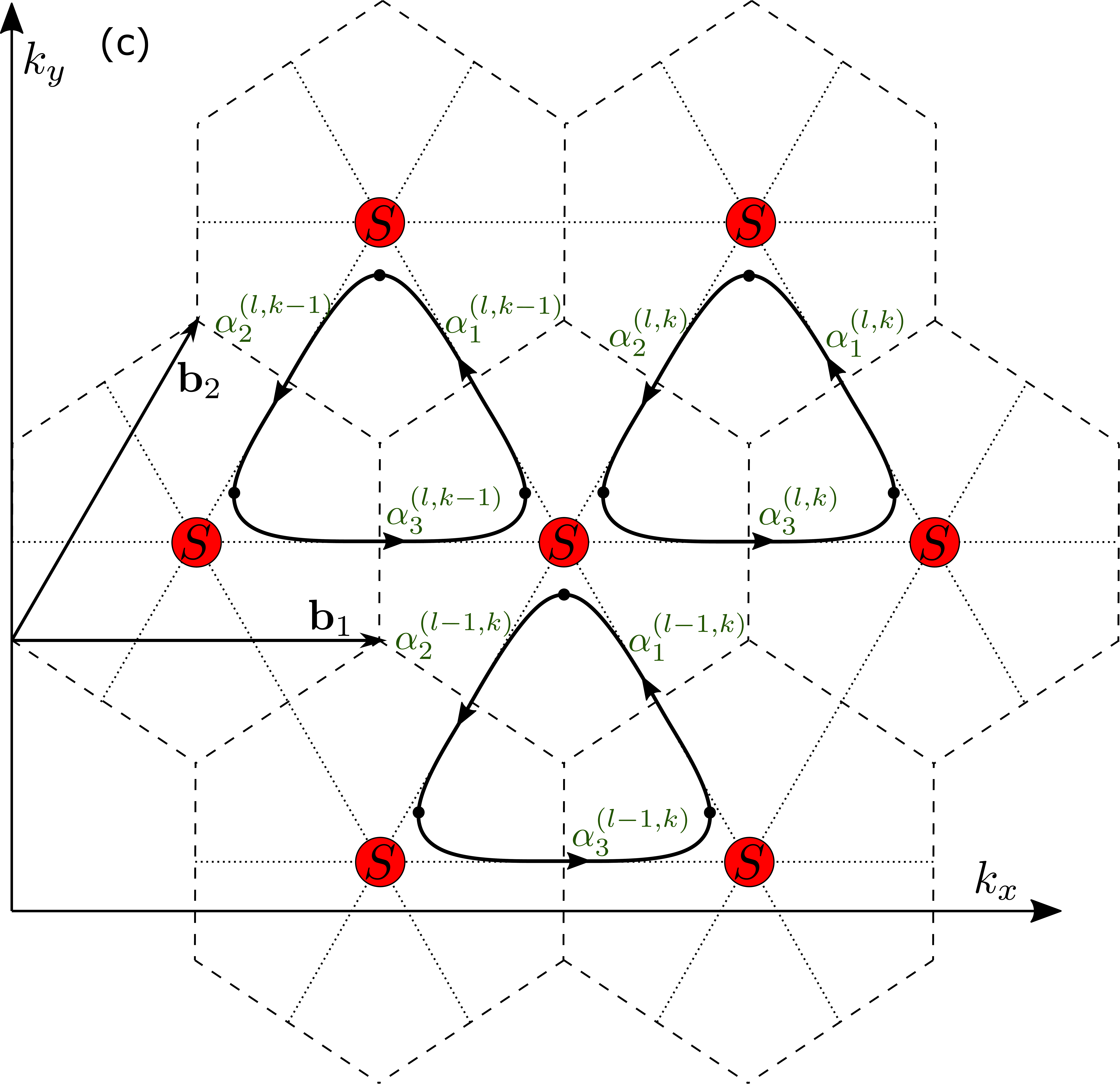}
		\caption{(a): Effective dispersion around Monkey saddle point in momentum space given by Eq.~\eqref{eq:Monkey-saddle} with three trajectories on a single energy level coming close at the MB region (shown as blue lines). (b): Tight-binding dispersion of triangular lattice with imaginary hoppings \eqref{eq:effective-dispersion-triangular-lattice} in which monkey saddle points connect cyclotron orbits into network. (c) Planar orbit network for dispersion (b) in rotated coordinate frame with semiclassical regions labeled by weight coefficients $\alpha$ [see Eq.~\eqref{ee:WKB-solution-ansatz}], and MB regions with s-matrix (red circles). Reciprocal lattice vectors $\vec{b}_i$ and highly-symmetric lines are shown.}
		\label{fig:monkey-triangular-fig-1}
	\end{figure}
	
	As was found in the 1960s in pioneering works by Pippard \cite{Pippard1962a, Pippard1964a, Pippard1965a}, and Chambers \cite{Chambers1965PhysRev, Chambers1966dHvA,Chambers1968Osc_magnetores,Chambers1973linear_network_osc}, and summarized in Ref.~\cite{Fischbeck1970}, MB can lead to formation of coherent orbit networks composed of localized Landau level states (LLs) connected via tunneling between them. For 2D materials, the orbit network occurs in the vicinity of energy levels where the Bloch band in momentum space has saddle points located at the boundaries of the Brillouin zone (BZ). Then, tunneling between orbits in different cells of the extended BZ scheme forms a network (see Fig.\ref{fig:monkey-triangular-fig-1}). In the real space, this corresponds to a network of semiclassical cyclotron orbits, which makes LLs to be extended \cite{Pippard1962a,Chambers1965PRA,Fischbeck1970}. The discovery of novel 2D materials \cite{Novoselov2004Science,Novoselov2005,Novoselov2016,Manzeli2017Review_TMDC} dramatically increased the number of lattice geometries in which orbit networks can be formed.  Below we calculate the detailed structure of these states as well as their band dispersion.  In addition, we show that such extended LLs allow for longitudinal bulk conductance in the quantum Hall bar, which strongly exceeds the standard edge conductance \cite{Girvin1990QHE}. 
	\begin{figure*}
		\centering 
		\includegraphics[scale=0.365, valign=t]{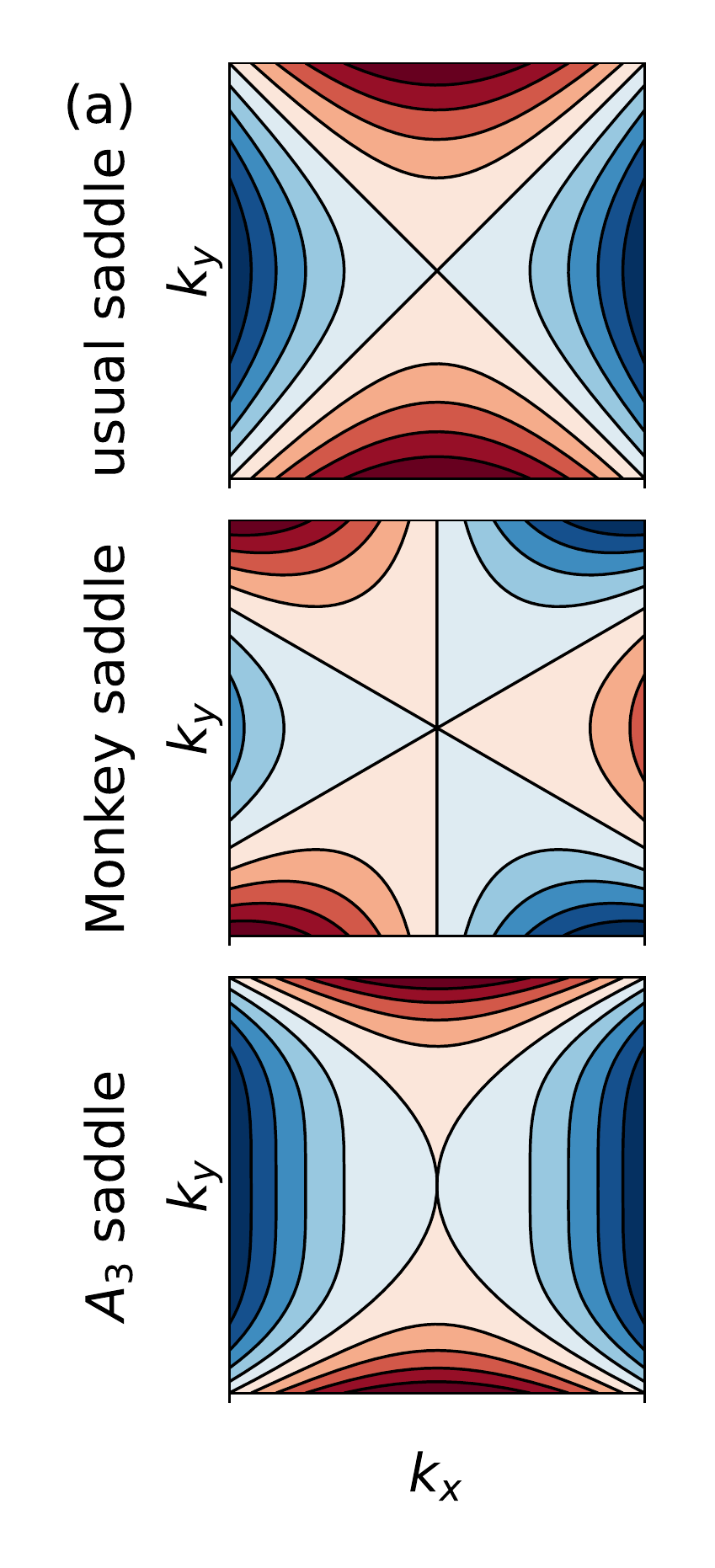}
		\includegraphics[scale=0.365, valign=t]{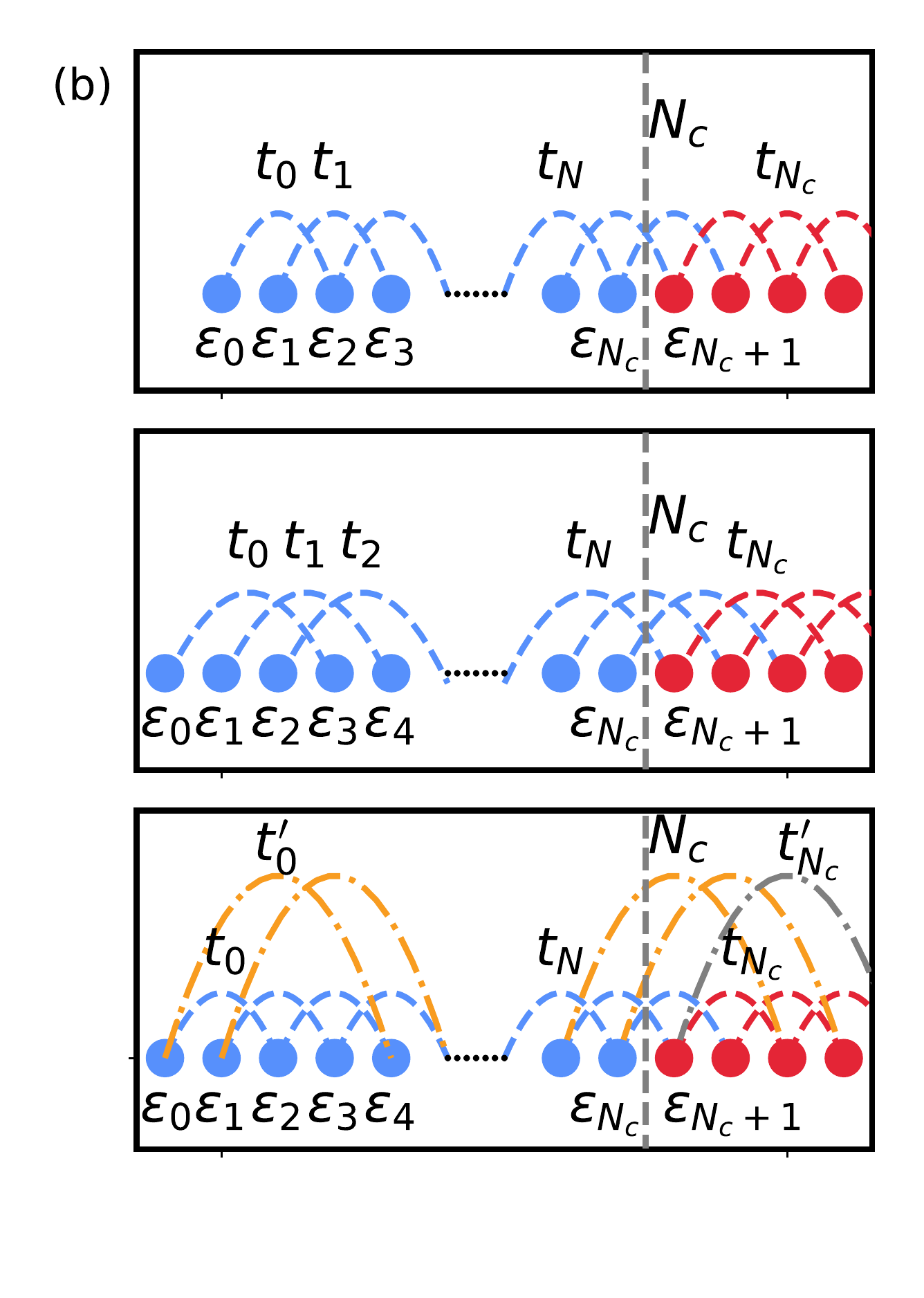}
		\includegraphics[scale=0.365, valign=t]{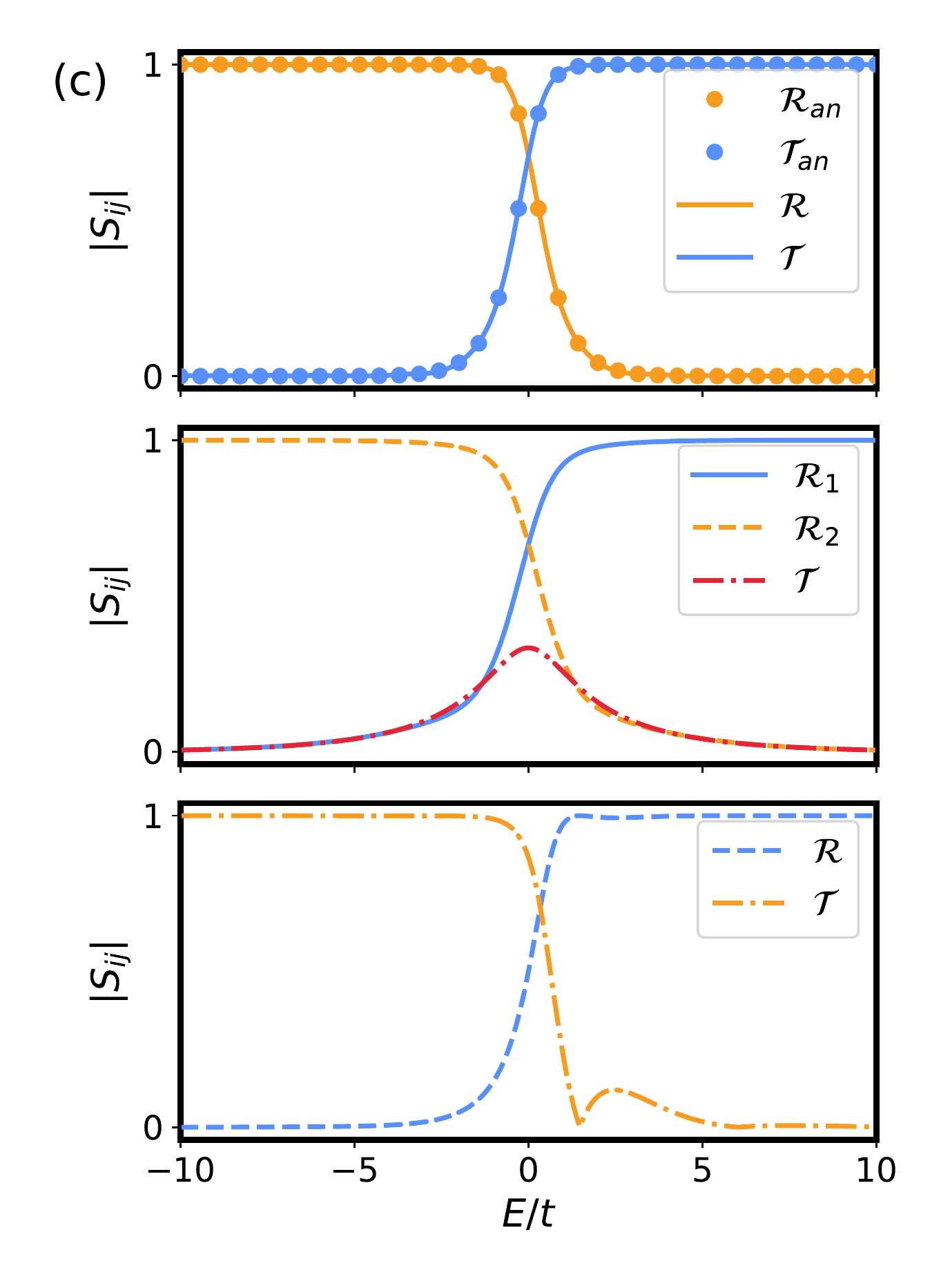}
		\includegraphics[scale=0.365, valign=t]{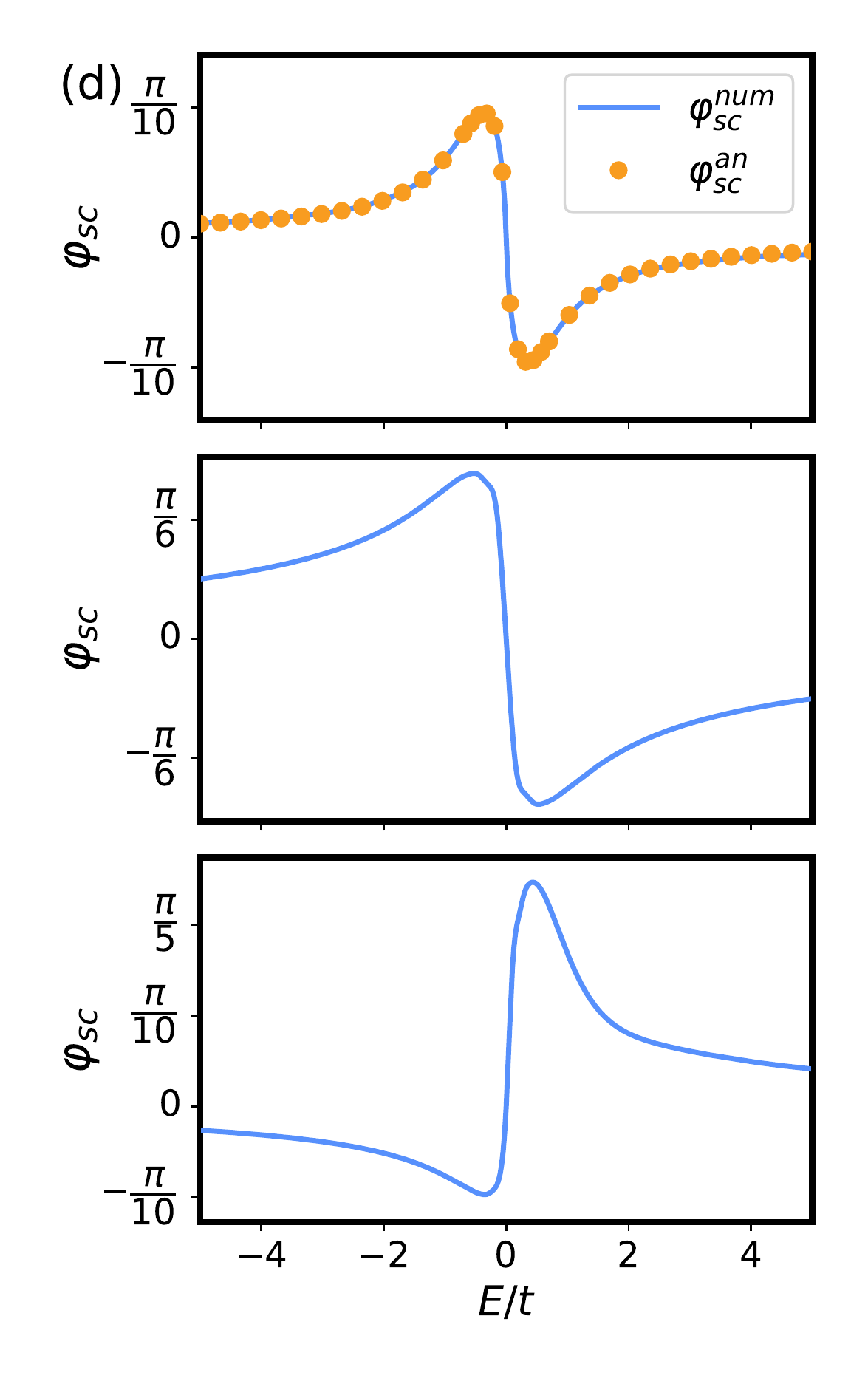}
		\caption{(a) The geometry of usual, Monkey and $A_3$ saddle points with equi-energy contours in $k-$space. (b): The corresponding 1D chains after mapping. The cut-off region with plane wave approximation is shown as a set of red sites with equal hopping parameters.  (c) The absolute value of s-matrix elements (reflection - $\mathcal{R}_i$ and transmission - $\mathcal{T}$). (d): The basis-independent  scattering phase calculated as $\arg(\det[S])$. For the usual saddle point (top panels in (c), (d)) we also show the comparison with exact analytic solution \cite{Azbel1961JETP,Azbel1964oscillations,Alexandradinata2018} marked by dots that perfectly agrees with the results obtained with our approach. In (c), (d) we set 
			$l_B=1$ and for any other magnetic field the results can be obtained by proper rescaling of energy \cite{footnote_rescaling}.}
		\label{fig:1D-mapping-fig-2}
	\end{figure*}

	There are two regimes of transport that orbit networks can govern: coherent regime with quantum phase that is accumulated along the cyclotron orbits and defines the exact energy spectrum, and incoherent regime with quantum phase averaged by the presence of disorder. Below, we describe the coherent regime and corresponding observable signatures that allow us to distinguish between different types of MB that happens at saddle points that connect cyclotron orbits. In addition, we note that the mini band structure appearing due to coherent orbit networks can be linked with the topological Hall effect of electrons in skyrmion crystals \cite{Goebel2017NJP}.

	Recent studies of coherent orbit networks in 1D geometry predicted a number of interesting effects such as magic zeros in Landau level spectra \cite{Fu2022magic_zeros} and broadening of the Landau levels by the coupling of Fermi arcs on opposite surfaces in Kramers-Weyl semimetals \cite{Lemut2020NPJ}. Also, the predicted spectrum by 2D incoherent orbit network shows relatively good agreement with the Hofstadter butterfly for twisted bilayer graphene \cite{Fertig2014}. The scaling of miniband width appearing from orbit networks with magnetic field was obtained for square lattice \cite{Gvozdikov2007PRB,Nikolaev2022} and graphene \cite{Nikolaev2021}.
	
	The semiclassical equations of motion for the electron in crystal under external weak magnetic field are given by the Lorentz force \cite{Pippard1960RPP,Lifshitz1960}
	\begin{align}\label{eq:equation-of-motion}
		\hbar\p_t\vec{k}=-c^{-1}e(\vec{v}_{k}\times\vec{B}).
	\end{align}
	with usual velocity replaced by group velocity found from the dispersion law $\vec{v}_{k}=\frac{1}{\hbar}\p_{\vec{k}}E(\vec{k})$ that depends on wave vector $\vec{k}$. Here, we consider 2D crystals placed in perpendicular magnetic field along the $z$-direction $\vec{B}=(0,0,B)$. Equation~\eqref{eq:equation-of-motion} restricts quasiparticles to move only along the lines of constant energy in momentum space. In Fig.~\ref{fig:monkey-triangular-fig-1} such lines are shown in the vicinity of the monkey saddle point and in the dispersion of the tight-binding model introduced below in Eq.\eqref{eq:effective-dispersion-triangular-lattice}. Before proceeding to MB, we note that after integration over time in Eq.~\eqref{eq:equation-of-motion} one finds that the trajectory in real space is rotated by angle $\pi/2$ compared to the $E(\vec{k})=\text{const}$ line in $k$-space and its size is rescaled by squared magnetic length \cite{Lifshitz1960,Stark1967}:
	\begin{align}
		k_x=-\frac{y-y_0}{l_B^2}, \,\, k_y=\frac{x-x_0}{l_B^2},\,\, l_B=\sqrt{\frac{\hbar c}{e B}}.
	\end{align}
	In what follows, we set $\hbar=c=1$. Magnetic field is considered as weak if magnetic length is much larger than the lattice constant of the crystal, $l_b\gg a$.
	
	We now focus on the detailed description of tunneling that takes place in the vicinity saddle point in dispersion due to magnetic breakdown. The saddle points are defined as points where the gradient of dispersion vanishes, $\nabla_{\vec{k}}E(\vec{k})=0$. As was shown in Ref.~\cite{Yuan2019Nature}, they can be further classified as usual or high-order depending on the ``flatness'' of dispersion around that point. More formally, the usual type corresponds to non-vanishing determinant of Hessian matrix $\mathcal{D}_{ij}=\p_{k_i}\p_{k_j}E(\vec{k})$ for dispersion, while the high-order ones have zero determinant and optionally zero Hessian itself. They could be further classified into many types depending on the underlying symmetry point group, see Refs.~\cite{Chamon2020,Yuan2020PRB}. Below, we show that our approach works for all possible saddle points. The magnetic breakdown happens because several constant energy lines in $k$-space come close to each other near the saddle point, see Fig.~\ref{fig:monkey-triangular-fig-1}(a). Thus, the tunneling probability between them becomes of order of one, and therefore we have to properly solve the scattering problem in the corresponding region. The complication arises due to the fact that typical dispersion around the saddle point has high powers of polynomials in $\vec{k}$, for example 
	\begin{align}\label{eq:Monkey-saddle}
		E_M(\vec{k})=-t a^3 (k_x^3-3 k_x k_y^2)
	\end{align} 
	for the monkey saddle. Here $t$ is a constant with dimension of energy. Generally, it is not possible to solve a Schr\"{o}dinger equation for such a Hamiltonian analytically to match it with plane-wave solutions away from the MB region. The only available closed form solution of such kind exists for the usual saddle point \cite{Azbel1961JETP,Alexandradinata2018}, a partial case of $A_{2n-1}$ points, 
	\begin{align}\label{eq:usual-saddle-A3}
		E_{log}(\vec{k})=ta^2(k_x^2-k_y^2),\,\,E_{A_{2n-1}}(\vec{k})=t(a^2 k_x^2-(ak_y)^{2n}),
	\end{align}
	with $n=1,2,...$.
	But we show that our semi-numerical method efficiently solves the Schr\"{o}dinger equation up to any precision and enables us to find the s-matrix.
	
	To introduce our method, we use the cylindrical gauge for vector potential $\boldsymbol{A}=\frac{B}{2}(-y, x, 0)$ and make use of the oscillator-type basis for Landau levels $\lvert n\rangle$, with their coordinate representation given by:
	\begin{align}\label{eq:oscillator-basis}
		\psi_n(x, y)=\left(\frac{\partial}{\partial w}-\frac{w^*}{4 l_B^2}\right)^n w^n e^{-|w|^2 / 4 l_B^2},\quad w=x+i y.
	\end{align}
	Using Landau level basis~\cite{Footnote1}, the effective Hamiltonian of the saddle point in magnetic field that is written in terms of canonical momenta $\Pi_i=k_i+e A_i$ can be expressed in terms of ladder operators by using the replacement:
	\begin{align}\label{eq:canonical-momenta}
		k_x\to\Pi_{x}=\frac{\hat{a}+ \hat{a}^{\dagger}}{\sqrt{2} l_B},\,k_y\to\Pi_{y}=\frac{i(\hat{a}- \hat{a}^{\dagger})}{\sqrt{2} l_B},
	\end{align}
	with standard commutation relation $\left[\hat{a}, \hat{a}^{\dagger}\right]=1$.
	In the simplest case of the usual saddle point, we find 
	\begin{align}
		& H_{\log }=t a^2 l_B^{-2} [\hat{a}^2+\left(\hat{a}^{\dagger}\right)^2].
	\end{align}
	Here, we rescaled energy by $t$ and set $l_B = a = 1$, which can be later restored by rescaling energy dependence of the s-matrix. The more complicated example of monkey saddle \eqref{eq:Monkey-saddle} with mixed $k_x k_y^2$ product requires a symmetrization procedure to make the Hamiltonian Hermitian in terms of ladder operators. In the general case, different symmetrizations of particular polynomial Hamiltonian give different results for the lower order terms due to non-trivial commutation relations. To uniquely fix the symmetrization procedure, we expand the tight-binding Hamiltonian of the underlying lattice with assumption that momenta operators do not commute. For the Monkey saddle after simplification this reads [see Appx.\ref{app:mb}]
	\begin{align}
		&H_M= -\frac{t a^3 l_B^{-3}}{2\sqrt{2}}\left[\left(\hat{a}+\hat{a}^{\dagger}\right)^3+3\left(\hat{a}-\hat{a}^{\dagger}\right)\left(\hat{a}+\hat{a}^{\dagger}\right)\left(\hat{a}-\hat{a}^{\dagger}\right)\right].
	\end{align}
	We note that in more general case of higher polynomial Hamiltonians one might find different symmetrization results depending on the lattice. If the tight-binding Hamiltonian is not known exactly, all possible symmetrizations that give different expressions in terms of ladder operators should be analyzed.
	
	Next, we explain how to obtain the scattering matrix that describes magnetic breakdown around a saddle point.
	We start by noting that the exact solution of the Sch\"{o}dinger equation $H\Psi=E\Psi$ with $\Psi=\sum_{n=0}^{\infty} \phi_n|n\rangle$ yields a set of recursive equations.
	For a usual van Hove singularity, we find  
	\begin{align}
		&E \varphi_0-\sqrt{2} \phi_2=0,\quad E \phi_1-\sqrt{6} \phi_3=0,\nn
		&E \phi_n-\sqrt{n(n-1)} \phi_{n-2}-\sqrt{(n+1)(n+2)} \phi_{n+2}=0.
	\end{align}
	Recursive equations for other saddle points are derived in the Appendix \ref{app:mb}.
	We note that a set of recursive equations can be mapped onto a 1D tight-binding problem: the term multiplying $\varphi_n$ corresponds to an on-site potential for the site with index $n$, while the terms involving $\varphi_m$ with $m\neq n$ represent the tight-binding hopping parameters that connect the $n$-th site to the $m$-th site.
	By imposing truncation at large index $n=N_c$ and replacing all remaining equation with those where $n=N_c$, we obtain a natural mapping to $N_c$-site 1D chain of atoms connected to a translationally invariant semi-infinite lead, shown in Fig.\ref{fig:1D-mapping-fig-2}(b).
	Then, we obtain the s-matrix using the propagating modes of the lead at energy $E$, with the number of scattering states corresponding to the number of semiclassical orbits coming close at the MB region.
	
	However, the obtained s-matrix is in the LL basis.
	To transform the s-matrix into basis of modes with a definite angle in momentum space, we use the creation ladder operator
	\begin{align}\label{eq:a-dag-kxy-relation}
		a^{\dagger}\sim k_x+i k_y\equiv k e^{i\phi_k},
	\end{align}
	where $\phi_k$ is the angle in momentum space. 
	Hence, performing a basis transformation on the propagating modes in semi-infinite leads to a basis where $a^\dagger$ is diagonal converts the obtained s-matrix into a physical one. 
	The technical details of this procedure for the usual and Monkey saddle are discussed in the Appendix \ref{app:mb}. The chirality and consequent absence of backscattering of the states with definite angle, that are spatially separated, ensures the unique definition of the physical s-matrix.
	
	For some saddle points the asymptotic modes at large momenta are indistinguishable by their angle in momentum space. In this case, we cannot apply our procedure of transforming the s-matrix into a physical basis. An example of such a saddle point is $A_{2n-1}$ described by Eq.~\eqref{eq:usual-saddle-A3}. For this saddle point, the angle of trajectory in momentum space with respect to the $x$-axis tends to zero as the wave number tends to infinity, see the bottom panel of Fig.\ref{fig:1D-mapping-fig-2}(a).
	We resolve this by introducing angle-fixing regularization, achieved through the inclusion of sufficient amount of sub-leading terms in the effective Hamiltonian
	\begin{align}
		E_{A_{2n-1}}^{\prime}(\boldsymbol{k})=t\left(a^2 k_x^2-a^{2n} (k_y^{2n} - \beta k_x^{2n})\right),\,\,n\geq 2,
	\end{align}
	where we use the $\beta>0$ constant as a regularization parameter and this parameter defines angles far away from the scattering region, not playing a role in the vicinity of the saddle point.
	We choose the truncation number $N_c$ such that the leading terms strongly dominate in effective 1D tight-binding equations and the mode separation into the angle basis can be done with good precision: $t_{N_c+1}^{max}\gg E,...$.
	Physically, this corresponds to taking the region where the scattering between modes with different angles is absent.
	
	To demonstrate our method, we numerically solve for the scattering matrix using Kwant code \cite{Kwant_paper2014,codereference}.
	We show our results for the absolute values of the transmission and reflection elements of the s-matrix and scattering phases in panels (c) and (d) of Fig.\ref{fig:1D-mapping-fig-2}.
	All these elements are gauge-invariant and independent of incoming and outgoing basis modes selection. In the case of usual van Hove singularity, it demonstrates perfect agreement with analytic expressions [see \cite{Alexandradinata2018}, Appx. \ref{app:on}]. 
	For the $A_3$ saddle point, we find a nontrivial behavior of transmission coefficients shown in bottom row in Fig.\ref{fig:1D-mapping-fig-2}(c). The presence of zeros in the transmission coefficient signifies the complete reflection of a quasiparticle moving along a cyclotron trajectory at that specific energy. Consequently, this phenomenon results in the effective reduction of orbit network to a single cell. The manifestation of this effect is demonstrated below by the narrowing of the mini-band width in the spectrum and the corresponding reduction in bulk conductance.  
	
	\begin{figure*}
		\centering
		\includegraphics[scale=0.35]{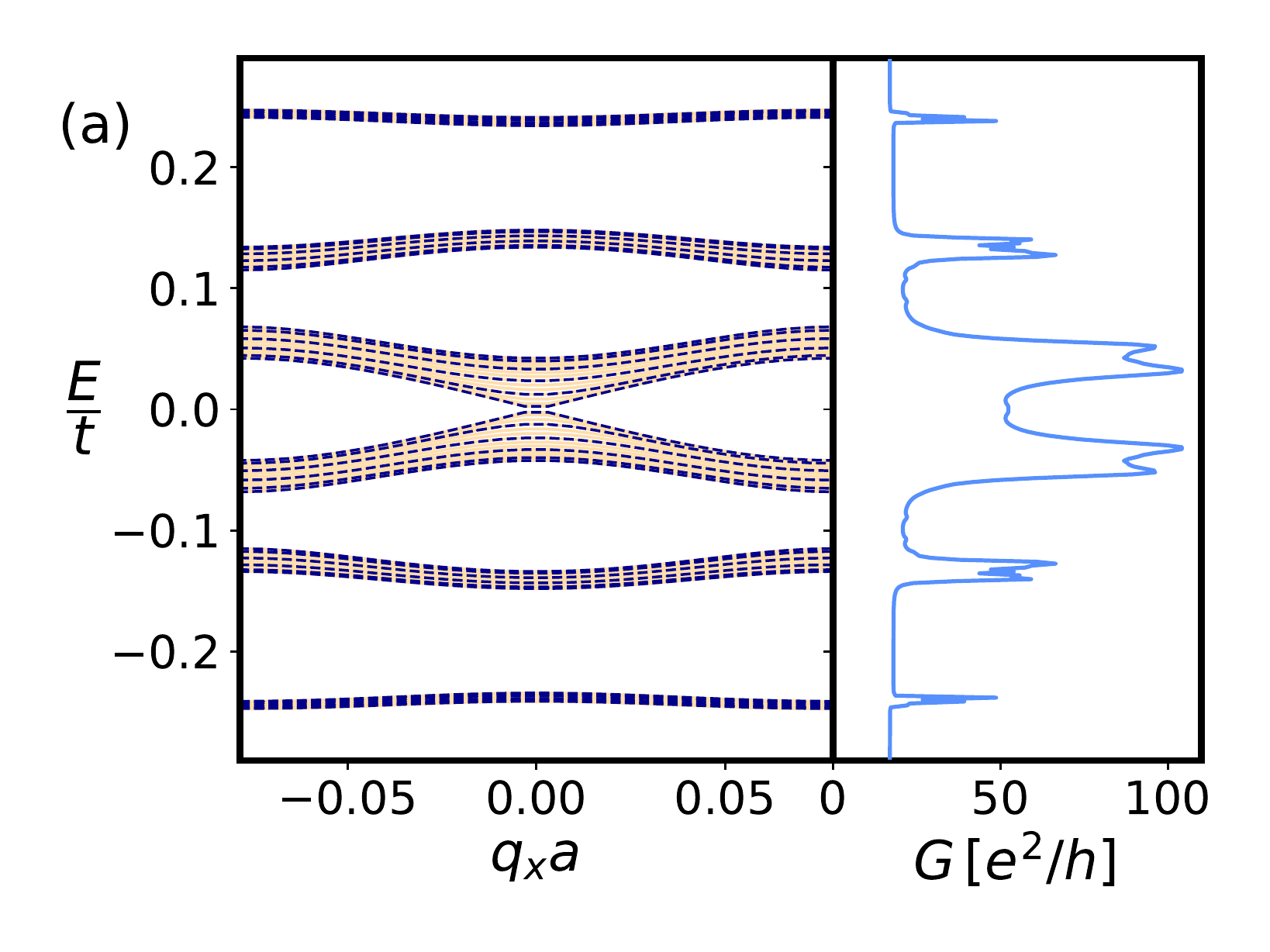}
		\includegraphics[scale=0.35]{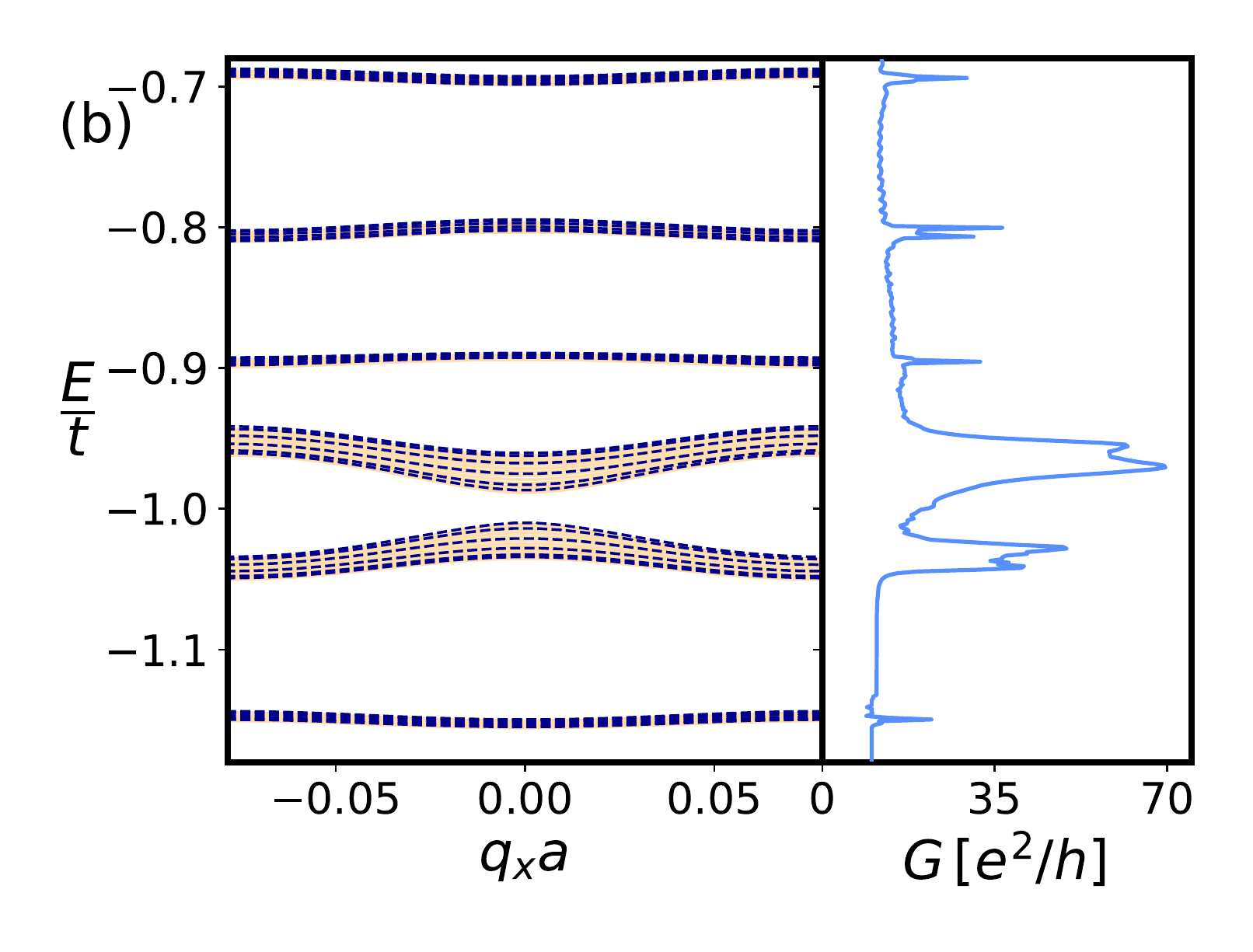}
		\includegraphics[scale=0.35]{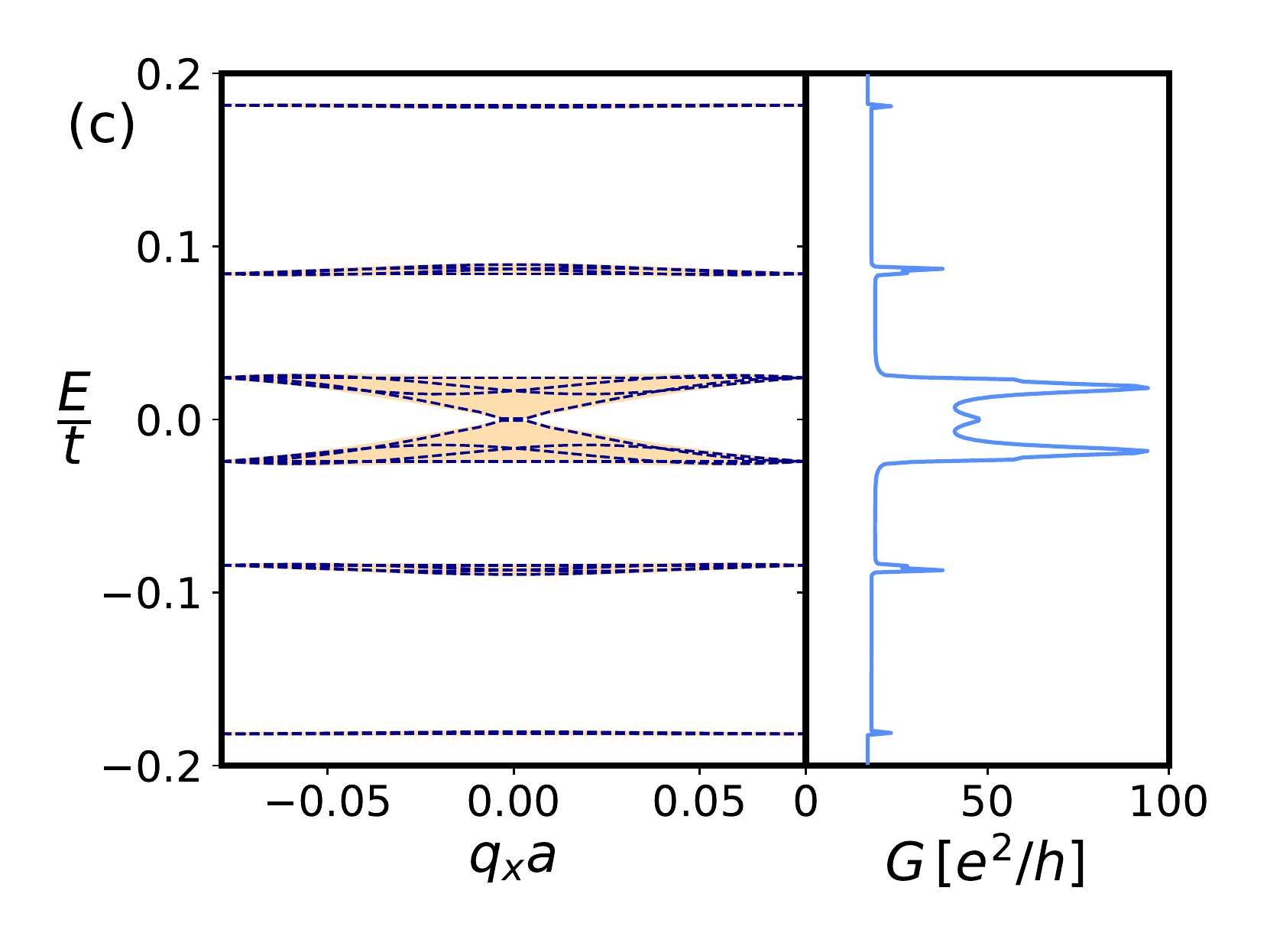}
		\caption{Comparison of Landau mini-band spectrum with longitudinal conductance in the quantum Hall bar for three systems: (a), (b) square lattice with usual (hopping $t_2=0$ in Eq.~\eqref{eq:square-lattice-NN-3NN-Hamiltonian}) and $A_3$ ($t_2=t/4$) saddle points, and (c) triangular lattice with imaginary hoppings that contain Monkey saddles. LL mini-bands (orange solid lines) obtained from tight-binding simulations are compared with solutions of spectral equations \eqref{eq:spectral-eq-square-lattice},\eqref{eq:spectral-eq-triangular-lattice} for a set of $q_y a$ (blue dashed lines). The flux value per unit cell was taken equal to $\Phi=1/40\,\Phi_0$ and the width of the Hall bar was $W=320a$. For the spectrum calculation a periodic boundary condition was imposed. A single period of Landau mini-band oscillation is shown and corresponds to $\sqrt{\Phi/\Phi_0}$ part of BZ. The width of conductance peaks measures the Landau mini-band broadening. In panel (b) the miniband around $E\approx -0.8 t$ is wider than the one at $E\approx -0.9 t$ due to the first zero of transmission coefficient at the $A_3$ saddle point (see Fig.~\ref{fig:1D-mapping-fig-2}(c), \cite{footnote_rescaling}). }
		\label{fig:square-lattice-all-fig-3}
	\end{figure*}
	With the complete description of MB at hand, we now propose a transport setup which would probe the features of the high-order saddle points. Since our goal is to distinguish energy dependence of both scattering amplitude and phase of MB at different saddle points, we use the coherent orbit networks that appear when the saddle points are placed at the edge of the BZ. Such coherent orbit networks were widely discussed in literature in the late 1960s \cite{Pippard1962a,Chambers1965PRA,Fischbeck1970}, but the absence of experiments with 2D atomically-thin crystals limited discussion to the simplest geometries, such as weak perturbative potential with square lattice geometry. Here, we use the same approach of magnetic translation symmetry groups and describe the orbit networks that are connected via usual as well as high-order saddle points. As it is clear from the constant energy curves in the spectrum in the extended BZ scheme (see Fig.~\ref{fig:monkey-triangular-fig-1}(b)), the orbits networks in $k$ space have perfect periodicity and thus should be periodic in $r$ space. However,
	in the presence of external magnetic field, the translation operators of the lattice $\hat{T}_{\vec{R}_i}=\exp\{\vec{\nabla}_r\vec{R}_i\}$, with $\vec{R}_i$ being a basis vector, should be replaced by magnetic translation operators \cite{Fischbeck1970}, which up to phase factor are equal to 
	$
	\hat{T}_{\vec{R}_i}^{M}=\exp\{\left(\vec{\nabla}_r+ie(\vec{A}(\vec{R}_i)+\vec{R}_i\times\vec{B})\right)\vec{R}_i\}.
	$
	The corresponding operators define a magnetic unit cell. To obtain a closed set of equations for the orbit network, we should restrict the value of magnetic flux per unit cell of the lattice to be a rational number
	\begin{align}
		\Phi=B\left|\mathbf{R}_1 \times \mathbf{R}_2\right|=B \frac{(2 \pi)^2}{\left|\mathbf{b}_1 \times \mathbf{b}_2\right|}=\frac{q}{p} \Phi_0,\,\, \Phi_0=\frac{h}{e}.
	\end{align}
	Here, $\vec{b}_i$ are the basis vectors of reciprocal lattice. In the further calculations, we restrict ourselves to the case of $q=1$.
	This relation is equivalent to setting magnetic unit cell to the integer number $p$ of lattice unit cells. Now we are ready to proceed with defining a basis of semi-classical wave functions on the links of networks. 
	These are 
	Zilberman-Fischbeck (ZF) wave functions \cite{Zilberman1957,Fischbeck1970,Alexandradinata2018}, written using the WKB-type approximation far from scattering region. The ZF functions are expressed in a  gauge-invariant coordinate space with replacement $\Pi_x \to k_x$ and $\Pi_y \to -i l_B^{-2}\p_{k_x}$. Since the scope of this paper is limited by the linear effects in magnetic field, we use the first order expansion of ZF functions with $a^2/l_B^2\ll1$:
	\begin{align}\label{eq:ZF-WKB-func}
		\Psi_{ZF}(k_x)=\left|\frac{\partial E(\vec{k})}{\partial k_y(k_x)}\right|^{-\frac{1}{2}}\exp\left[-i l_B^2 \int_{k_{x,0}}^{k_x}k_y^{E}(k_x)\,d k_x\right].
	\end{align}
	Here $k_y^E(k_x)$ stands for the solution of constant energy contour equation $E(k_x,k_y^E(k_x))=E$.
	The full wave function of the orbit network state is composed as a weighted superposition of the $\Psi_{ZF}$ wave functions in different unit cells:
	\begin{align}\label{ee:WKB-solution-ansatz}
		\Psi(k_x)&=\sum_{k, l}^{\infty} e^{i l_B^2 [k_x l b_{2,y}  -\frac{l^2}{2}b_{2,x}b_{2,y}]} \nn
		&\times\sum_{j} \alpha_j^{(l, k)} \Psi_{ZF}^j\left(k_x-k b_{1,x}-l b_{2,x} \right).
	\end{align}
	In this expression, each weight coefficient $\alpha_j^{(l, k)}$ contains two cell indices $l,k$ as well as a unique index $j$ corresponding to the different parts of the orbit between scattering points inside single cell of the network. The example of this notation is shown in Fig.~\ref{fig:monkey-triangular-fig-1}. Due to periodicity of the network, the solutions have the form of Bloch waves $\alpha^{(l, k)}_{j}=\alpha_j e^{i\left(p_k k+p_l l\right)}$. 
	The magnetic translation group restricts the allowed values of
	$p_{l,k}$ to particular dependence on translation operator $\hat{T}_{\boldsymbol{R}_i}^M$ eigenvalues $\vec{q}$:  $p_l=-l_B^2\left(q_x b_{2, y}-q_y b_{2, x}\right)$, $p_k=l_B^2 q_y K_{1, x}$ [see Appx. \ref{app:on}].
	Next, we use the S-matrices obtained above to couple the ZF solutions in the neighboring cells. By noting that ZF functions from Eq.~\eqref{eq:ZF-WKB-func} correspond to the modes with proper angles, we can straightforwardly insert parameters of the s-matrix into the system of equations, and write it in the form of a Ho-Chalker operator \cite{HoChalker1996PRB}:
	\begin{align}\label{eq:Ho-Chalker-general}
		\hat{S}_{HC}(E,\vec{q})\vec{\alpha}=0.
	\end{align} 
	While substituting the s-matrix,  
	we subtracted the difference in dynamical phases of modes with defined angles and ZF functions \eqref{eq:ZF-WKB-func} at given energy.
	Such a difference appears due to the fact that in geometry of the scattering problem 
	one assumes semiclassical ZF solutions with the phase fixed at infinity, while in the  
	orbit network ZF function phase is fixed at particular point inside the network unit cell.  
	
	The nonlinear eigenvalue problem for the Ho-Chalker operator \eqref{eq:Ho-Chalker-general} can be rewritten in the form of spectral equation $\operatorname{det} \hat{S}_{H S}(E, \mathbf{q})=0$ for a given lattice model [see Appx. \ref{app:on}]. Below we demonstrate this for square and triangular lattice, and show that the MB s-matrix calculated above plays a key role in definition of the properties of coherent orbit network.
	In the case of square lattice with only first and third NN hoppings taken into account,
	\begin{align}\label{eq:square-lattice-NN-3NN-Hamiltonian}
		H_{sq}(\vec{k})=-2 \sum\limits_{i=x,y}\left(t \cos k_i a+ t_3\cos 2 k_i a\right),
	\end{align}
	the spectral equation is:
	\begin{align}\label{eq:spectral-eq-square-lattice}
		&\cos \left(\frac{l_B^2\mathcal{A}(E)}{2}-\varphi_{sc}\right)=\nn
		&
		\pm \mathcal{T}\mathcal{R} \left[\cos \left(l_B^2 q_1 b_{2, y}\right)+\cos \left(l_B^2 q_2 b_{1, x}\right)\right].
	\end{align}
	Here, $\mathcal{A}(E)$ is the area enclosed by the constant energy curve in momentum space. The elements of the s-matrix, denoted as $\mathcal{R}$-reflection, $\mathcal{T}$-transmission and $\varphi_{sc}=\arg(\det S)$ is the scattering phase, are shown in Fig.~\ref{fig:1D-mapping-fig-2}. For such a lattice Hamiltonian, the connection of orbits happens via usual van Hove singularity at the X-point of BZ for $t_3=0$ or via high-order van Hove singularity of $A_3$ type for $t_3=t/4$.
	In the case of the triangular lattice with imaginary hoppings, the dispersion is
	\begin{align}\label{eq:effective-dispersion-triangular-lattice}
		&H_{tr}(\mathbf{k})=\nn
		&2 t \left(\sin k_x a- \sin \frac{ k_x -\sqrt{3} k_y}{2}a - \sin \frac{k_x+\sqrt{3} k_y}{2}a\right).
	\end{align}
	The monkey saddle (see Fig.~\ref{fig:monkey-triangular-fig-1}) connects orbits from different cells into a network. The corresponding spectral equation is
	\begin{align}\label{eq:spectral-eq-triangular-lattice}
		&\cos \left(\frac{l_B^2 \mathcal{A}(E)-\varphi_{s c}}{2}\right)=\mathcal{T}\left[\cos \left(l_B^2 q_2 b_{1, x}-\frac{\pi p}{2}\right)\right.\nn
		&+\left. \cos \left(l_B^2\left[q_1 b_{2, y}+q_2\left(b_{1, x}-b_{2, x}\right)\right]-\frac{\pi p}{2}\right)\right.\nn 
		&+\left.\cos \left(l_B^2\left[q_1 b_{2, y}-q_2 b_{2, x}\right]-\frac{\pi p}{2} \right)\right].
	\end{align}
	The left-hand side of each spectral equation, as defined in \eqref{eq:spectral-eq-square-lattice} and \eqref{eq:spectral-eq-triangular-lattice}, yields the conventional flat Landau levels when equated to zero. On the other hand, the nonzero right-hand side converts Landau levels into minibands. The width of these minibands is determined by the van Hove singularity, the s-matrix transmission coefficient, and the lattice-specific dispersion. To explore this behavior, we numerically solve \cite{codereference} the spectral equations for different values of $q_x$ and for a small set of $q_y$. The resulting miniband structures are depicted by the blue dashed lines in Fig.~\ref{fig:square-lattice-all-fig-3}, showing both the width and internal structure of analytic spectrum of a mini-bands. The spectrum obtained from a tight-binding simulations \cite{codereference,Kwant_paper2014} is shown as orange lines filling the corresponding regions and demonstrates excellent agreement with the semi-classical predictions. For our analysis, we utilized a narrow Hall bar geometry with periodic boundary conditions, having a width several times larger than the magnetic unit cell. That width is already enough to have many bulk conducting states inside the orbit network.

	The appearance of oscillating dispersion and broadening of Landau levels due to orbit networks is expected to be manifested in the transport experiments such as Shubnikov-de-Haas oscillations or high-frequency magnetic breakdown oscillations \cite{Gerhardts1989,Winkler1989,Beenakker1989,Steda1990,Gvozdikov2007,Lemut2020NPJ}. As the most pronounced signature, we present a calculation of longitudinal conductance in two-terminal Hall bar geometry. Typically, such conductance is governed by edge states \cite{Girvin1990QHE} and is strongly suppressed. As it is shown in the right part of each panel in Fig.~\ref{fig:square-lattice-all-fig-3}, the dispersive Landau mini-bands induce bulk conductance that is much larger than background edge conductance. We compared the spectrum for the lattices with periodic boundary condition with the conductance shape in finite size systems for the same values of magnetic field. The width of the peaks in the conductance agrees with the broadening of Landau mini-bands, thus providing a tool for estimation of the tunneling probabilities $\mathcal{T}$ for MB s-matrix at the saddle point. In addition, we note that the specific property of the $A_3$ saddle point with zero transmission coefficient [see Fig.\ref{fig:1D-mapping-fig-2}(c)] leads to a much smaller conductance peak at corresponding chemical potential comparing to other peaks, shown in Fig.\ref{fig:square-lattice-all-fig-3}(b).
	
	To give an estimate of magnetic field required for the experiment, we use an estimate of magnetic length $l_B\approx 26 \,\text{nm}/\sqrt{B[\text{T}]}$ with typical experimental values of magnetic field $B\sim 10\,T$ \cite{Finney2022PNAS}, which gives $l_B\approx 10\, \text{nm}$. The broadening of Landau miniband becomes significant compared to the hopping parameter (see Fig.~\ref{fig:square-lattice-all-fig-3}) and larger than disorder broadening for magnetic fluxes around $\Phi=10^{-2}\Phi_0$ per lattice unit cell. Thus, it requires lattice constant to be of the order of $a\sim l_B \sqrt{2\pi \Phi/\Phi_0} \sim 2.5  \text{nm}$. 
	Such an estimate shows that one requires extremely high magnetic field to measure such effects in conventional systems, such as highly doped monolayer graphene \cite{Rosenzweig2020ORL_overdoped_graphene}. But, such lattice constants are typical for the modern artificial lattices \cite{Slot2017} as well as for Moir\'{e} materials such as twisted bilayer graphene \cite{Kim2017,Cao2018,Kim2021ACSNano}. In addition, we point out that the method of solving the MB problem developed above can be applied for the systems with spin-orbit coupling such as  Moir\'e bilayer transition-metal dichalcogenides \cite{Hsu2021PRB}.
	The structure of the orbit network might be visualized by injecting the current at proper chemical potential level into the system via narrow lead. The picture of current density distribution 
	is expected to follow the pattern of orbit network shown in Fig.~\ref{fig:monkey-triangular-fig-1} and might be probed by STM-type microscopy techniques \cite{Andrei2012Exp,Xie2019Exp}.
	
	We are grateful to Carlo Beenakker, Gal Lemut, Jakub Tworzydlo, Mikhail Katsnelson, Michal Pacholski and Johanna Zijderveld for fruitful discussions. A part of code is written for the Kwant package \cite{Kwant_paper2014} by J.B. Weston and T.\"{O}. Rosdahl. V.A.Z. and D.O.O. acknowledge the support from the Netherlands Organization for Scientific Research (NWO/OCW) and from the European Research Council (ERC) under the European Union's Horizon 2020 research and innovation programme. A.M.B. acknowledges NWO (HOTNANO) for the research funding. A.R.A. acknowledges the NWO VIDI Grant (016.Vidi.189.180).
	
	V.A.Z. performed the analytical calculations and numerical tight-binding simulations for orbit networks. A.M.B. performed numerical calculations for scattering matrices and contributed to tight-binding simulations. A.R.A. formulated the idea of ladder operator approach for s-matrix calculation. D.O.O. organized the workflow, wrote the manuscript and helped with analytical calculations and tight-binding simulations. All authors contributed to reviewing and editing the manuscript.

	\let\oldaddcontentsline\addcontentsline
	\renewcommand{\addcontentsline}[3]{}
	\bibliography{references}

\begin{thebibliography}{52}
\bibitem [{foo()}]{footnote_rescaling}%
  \BibitemOpen
  \bibinfo {note} {For the Hamiltonian written in gauge-invariant coordinates
  $H=A \Pi_x^n+B \Pi_x^s \Pi_y^m=A k_x^n+B k_x^s\left(-\frac{i}{l_B^2}
  \frac{\partial}{\partial k_x}\right)^m$ it is possible to introduce a change
  of variables $q_x=k_x l_B^{\frac{2 m}{m+n-s}}$, that would convert it to
  $l_B=1$ Hamiltonian with additional factor $H=l_{B}^{\frac{2 m
  n}{m+n-s}}\left[A q_x^n+B q_x^s\left(-i \frac{\partial}{\partial
  q_x}\right)^m\right]$. This factor should be used to obtain energy dependence
  of S-matrix for any magnetic field value.}\BibitemShut {Stop}%
\bibitem [{Foo()}]{Footnote1}%
  \BibitemOpen
  \bibinfo {note} {We note that we take Landau basis with $m_z=0$. For the
  problems in empty space different $m_z$ states are trivially degenerate. The
  problem at hand does not mix different $m_z$ states.}\BibitemShut {Stop}%
\end{thebibliography}
	\let\addcontentsline\oldaddcontentsline

	\clearpage
\appendix
\onecolumngrid
\tableofcontents

\section{Magnetic breakdown near usual and high-order saddle points \label{app:mb}}
In this section, we describe in detail the way how we obtain the scattering matrix for the quasiclassical wave functions of WKB type that approach the saddle point of any topology in 2D dispersion. After introducing the key notation of Landau level basis in axial gauge and oscillator basis, we describe the full algorithm on example of usual saddle point 
that leads to logarithmic van Hove singularity (vHs). For such vHs the analytic S-matrix is known from exact solution of Schr\"{o}dinger equation \cite{Azbel1961JETP,Azbel1964oscillations,Alexandradinata2018} and can be compared with the results of our calculation. Next, we extend this algorithm to a number of high-order saddle points that were discussed in classifications in Refs.~\cite{Chamon2020,Yuan2020PRB}. In addition to comparing technical subtleties of realizations and related physical effects, we also compare the results with the simplest quasiclassical calculations of tunneling probability. 

\subsection{Oscillator basis of Landau levels and formulation of problem in terms of ladder operators}
We use the axial gauge for vector potential $\vec{A}=\frac{B}{2}(-y, x,0)$ and insert this into the effective model of quasiparticle with dispersion $\epsilon(k_x, k_y)$, which is expressed in terms of canonical momenta with standard commutation relation
\begin{align}
	H=\epsilon(\Pi_x,\Pi_y),\quad \Pi_i = k_i +e A_i,\quad [\Pi_x,\Pi_y]=-i l_B^{-2}.
\end{align}
Next, we introduce the ladder operators, 
\begin{align}
	\Pi_x=\frac{1}{\sqrt{2} l_B}\left(\hat{a}+\hat{a}^{\dagger}\right), \quad \Pi_y=\frac{i}{\sqrt{2} l_B}\left(\hat{a}-\hat{a}^{\dagger}\right), \quad\left[\hat{a}, \hat{a}^{\dagger}\right]=1,
\end{align}
with $l_B=\sqrt{\frac{\hbar}{e B}}$ being the magnetic length.
These operators are analogous to the ladder operators for the quantum harmonic oscillator. The basis of corresponding number operator $a^{\dagger} a|n\rangle=n|n\rangle$ with integer Landau level index $n\geq 0$ can be used to represent any polynomial Hamiltonian $H=\epsilon(\Pi_x,\Pi_y)$ as a matrix. The eigenstates $|n\rangle$ in the coordinate basis are given by 
\begin{align}
	|n\rangle=\psi_n(x, y)=\left(\frac{\partial}{\partial w}-\frac{w^*}{4 l_B^2}\right)^n w^n e^{-|w|^2 / 4 l_B^2},\quad w=x+i y,
\end{align}
and the matrix elements of ladder operators are
\begin{align}\label{eq:action-of-ladder-operators}
	\langle n|\hat{a}| m\rangle=\sqrt{m} \delta_{n, m-1}, \quad\left\langle n\left|\hat{a}^{\dagger}\right| m\right\rangle=\sqrt{m+1} \delta_{n, m+1}.
\end{align}
Next, we use this notation to obtain matrix representation of the effective Hamiltonians near different saddle points.

\subsection{Magnetic breakdown S-matrix for the usual saddle point}
\label{sec:supplement-log-vHs}
\begin{figure*}
	\centering
	\includegraphics[scale=0.4]{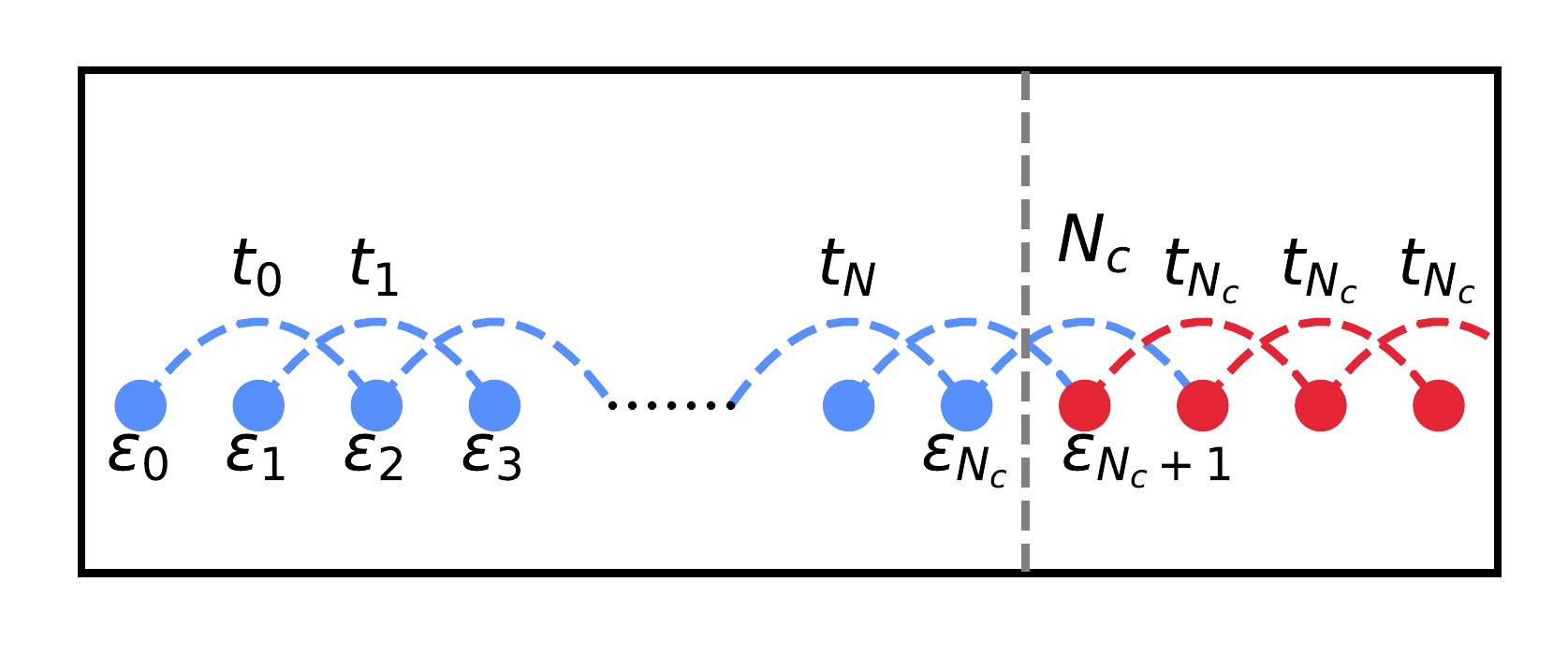}
	\caption{Effective 1D tight-binding model that represents the system of equations \eqref{eq:recursive-saddle-usual} with asymptotic truncation after Landau level index $n>N_c$.}
	\label{fig:saddle-point-ladder-chain}
\end{figure*}

The effective Hamiltonian in the vicinity of the usual saddle point in $k$-space of 2D band is given by
\begin{align}
	H_{\log}=\alpha k_x^2 - \beta k_y^2.
\end{align} 
Below, we set $\alpha=\beta=1$ for simplicity. Using the notation of ladder operators and oscillator basis, we rewrite this Hamiltonian in magnetic field as follows 
\begin{align}
	H_{\log }=\frac{1}{2 l_B^2}\left[\left(\hat{a}+\hat{a}^{\dagger}\right)^2+\left(\hat{a}-\hat{a}^{\dagger}\right)^2\right]=\frac{\hat{a}^2+\left(\hat{a}^{\dagger}\right)^2}{l_B^2}.
\end{align}
In the oscillator basis, this Hamiltonian is represented by the following matrix:
\begin{align}
	H_{l o g}=\frac{1}{l_B^2}\left(\begin{array}{cccccccc}
		0 & 0 & \sqrt{2} & 0 & 0 & 0 & 0 & \ldots \\
		0 & 0 & 0 & \sqrt{6} & 0 & 0 & 0 & \cdots \\
		\sqrt{2} & 0 & 0 & 0 & 2 \sqrt{3} & 0 & 0 & \ldots \\
		0 & \sqrt{6} & 0 & 0 & 0 & 2 \sqrt{5} & 0 & \ldots \\
		0 & 0 & 2 \sqrt{3} & 0 & 0 & 0 & \sqrt{30} & \ldots \\
		0 & 0 & 0 & 2 \sqrt{5} & 0 & 0 & 0 & \sqrt{N(N+1)} \\
		0 & 0 & 0 & 0 & \sqrt{30} & 0 & 0 & \ldots \\
		\ldots & \ldots & \ldots & \ldots & \ldots & \sqrt{N(N+1)} & \ldots & \ldots
	\end{array}\right).
\end{align} 
As a result, we transformed the problem into an eigenvalue equation, where the eigenstates are superpositions of oscillator basis states:
\begin{align}\label{eq:state-in-osc-basis}
	H\Psi = E\Psi,\quad \Psi=\sum_{n=0}^{\infty} \phi_n|n\rangle. 
\end{align}
We then reformulate this eigenvalue equation as a coupled set of recursive algebraic equations:
\begin{align}
	&E \phi_0-\sqrt{2} \phi_2=0, \nonumber\\
	&E \phi_1-\sqrt{6} \phi_3=0, \nonumber\\
	&E \phi_2-\sqrt{2} \phi_0-2 \sqrt{3} \phi_4=0, \nonumber\\
	&E \phi_3-\sqrt{6} \phi_1-2 \sqrt{5} \phi_5=0, \nonumber\\
	&\cdots \nonumber\\
	\label{eq:recursive-saddle-usual}
	&E \phi_n-\sqrt{n(n-1)} \phi_{n-2}-\sqrt{(n+1)(n+2)} \phi_{n+2}=0 .
\end{align}
The exact iterative solution of \eqref{eq:recursive-saddle-usual} quickly becomes very cumbersome because of the complexity of coefficients in the recursive relation. Thus, to solve this system of equations and those that shall appear for more complicated saddle points, we introduce a truncation scheme. 
After the large value of index $n>N_c$ we replace the index $n$ by $N_c$ in the coefficients of recursive relation. This procedure can be schematically illustrated by Fig.~\ref{fig:saddle-point-ladder-chain}, where we represent solution coefficients $\varphi_n$ as sites, coefficients of Hamiltonian that appear in recursive relations as hopping terms and onsite potentials, respectively. This correspondence allow us to solve the system of equations as an effective 1D tight-binding model of a chain. The introduced cut-off at large $N_c$ separates the scattering region from the translationally-invariant region with simplified coefficients. We note that there are two disconnected sets of equations, and this fact manifests the presence of two linearly-independent solutions in the original problem. 

After setting up the correspondence between effective Hamiltonian in the saddle point and 1D tight-binding model with two decoupled chains, we proceed with finding the S-matrix of magnetic breakdown. The procedure of truncation described above allows us to match the asymptotic solutions at high Landau level indices $n>N_c$ with exact solution at lower indices $n\leq N_c$. We solve the scattering problem numerically using the Kwant package for tight-binding simulations \cite{Kwant_paper2014}. However, the scattering modes, matched solutions and S-matrix obtained from numerical simulation are given in the oscillator basis. Thus, we perform an additonal procedure to link them with quasiclassical wave functions of cyclotron trajectories and corresponding magnetic breakdown S-matrix. To establish this link, we solve the problem analytically in the translationally-invariant region with $n>N_c$.

The eigenmodes of the infinite lead composed of two identical chains are given by two Bloch wave functions with identical degenerate band dispersions. These eigenmodes and their eigenenergies are given by the following expressions in the basis of two atoms per unit cells:
\begin{align}\label{eq:psi_modes_in_log_vhs_chain}
	\Psi_{1}(l)=\frac{1}{M}\sum_{n=N_c+1}^{\infty}e^{-2iln}|2n\rangle,\quad \Psi_{2}(l)=\frac{1}{M}\sum_{n=N_c+1}^{\infty}e^{-2iln}|2n+1\rangle,\quad \epsilon_{1,2}(l)=-E+2 t_{N_c} \cos 2 l .
\end{align}
Here, the length of the unit cell is $2$ because the coupling is between second neighbor oscillator basis states only. The normalization constant $M$ can be omitted in next calculations as it does not alter the final result. We note that the modes defined above are in the basis of oscillator states, with corresponding indices depicted as sites, and the momentum $l$ is defined in the corresponding reciprocal space. To obtain the S-matrix of physical modes, we have to establish connection between these modes and asymptotic modes far from the saddle point in $k$-space in the quasiclassical region (see Fig.~2(a) in the main text). To establish this connection, we notice that the modes in quasiclassical region are classified by their corresponding angle in momentum space: incoming modes correspond to angles $\pi/4$ and $5\pi/4$ for energies $E>0$, while outgoin modes are at $3\pi/4$ and $7 \pi / 4$ angles. 
We find the modes in angle basis by diagonalizing the creation ladder operator by noting that
\begin{align}
	\hat{a}^{\dagger}=\frac{l_B}{\sqrt{2}}(\Pi_x+i\Pi_y)\underset{B\to 0}{\to} k_x+i k_y = k e^{i\phi_k}.
\end{align}
The limit of zero magnetic field is used only to point out that the ladder operator $\hat{a}^{\dagger}$ allows one to classify propagating modes in lead according to the asymptotic angles of scattering modes in MB region that follow constant energy curves in saddle point dispersion. 
In other words, in the basis of eigenmodes of $a^{\dagger}$ operator, the phase of eigenvalue of $a^{\dagger}$ gives the angle of direction of propagation for incoming wave. Using the modes for the infinite lead defined in Eq.~\eqref{eq:psi_modes_in_log_vhs_chain} as a basis and taking into account that action of $a^{\dagger}$ on oscillator state shifts this state (see Eq.~\eqref{eq:action-of-ladder-operators}), we obtain another form of the operator $\hat{a}^{\dagger}$:
\begin{align}\label{eq:shift-operator-log-vhs}
	a^{\dagger}_{\Psi,in}=\begin{pmatrix}
		0 & e^{2 i l}\\
		1 & 0
	\end{pmatrix}.
\end{align}
This form of creation operator can be checked by direct action on basis states:
\begin{align}
	&\hat{a}^{\dagger}\Psi_{1}(l)=\frac{\sqrt{N_c+1}}{M} \sum_{N=N_c+1}^{\infty} e^{-2 i l N}|2 N+1\rangle\sim \Psi_2(l),\nn 
	&\hat{a}^{\dagger}\Psi_{2}(l)=\frac{\sqrt{N_c+1}}{M} \sum_{N=N_c+1}^{\infty} e^{-2 i l N}|2 N+2\rangle\sim e^{2 i k l}\Psi_1(l).
\end{align}
The eigenvalues and eigenvectors of $a^{\dagger}_{\Psi}$ operator are given by
\begin{align}\label{eq:ladder-operator-vhs-eigenvectors}
	\lambda_{1,2}=\pm e^{i l},\quad \chi_1=\frac{1}{\sqrt{2}}\begin{pmatrix}
		1\\
		e^{-i l}
	\end{pmatrix},\quad \chi_2 =\frac{1}{\sqrt{2}} \begin{pmatrix}
		-e^{i l}\\
		1
	\end{pmatrix}.
\end{align}
Notably, for momenta $l$ of the propagating modes close to Brillouin zone edge $l=\pm \pi/4$, we uncover the correspondence with asymptotic angles of the constant energy curves at large momenta that correspond to directions of incoming and outgoing modes in MB region.
Now we have to convert the $S_{\Psi}$-matrix from the basis of the modes \eqref{eq:psi_modes_in_log_vhs_chain} to the modes with definite angle. The S-matrix itself has the form 
\begin{align}
	S_{\Psi}=\begin{pmatrix}
		e^{i\alpha_1} & 0\\
		0 & e^{i\alpha_2}
	\end{pmatrix}
\end{align} 
with two phases calculated numerically by matching using Kwant code. The transformation is performed via rotation defined by eigenvectors in Eq.~\eqref{eq:ladder-operator-vhs-eigenvectors}, while taking into account that outgoing modes have opposite momenta $-l$ and the sign of momenta changes in $a^{\dagger}_{\Psi,out}$. For example, $a^{\dagger}_{\Psi,out}(l)=a^{\dagger}_{\Psi,in}(-l)$ in this system because of dispersion relation \eqref{eq:psi_modes_in_log_vhs_chain}. Then, the rotation to new basis of the S-matrix gives:
\begin{align}
	S=U_{out}S_{\Psi}U_{in}^{\dagger},\quad U_{out}=\frac{1}{\sqrt{2}}\begin{pmatrix}
		1 & - e^{-i l}\\
		e^{i l} & 1
	\end{pmatrix},\quad U_{in}=\frac{1}{\sqrt{2}}\begin{pmatrix}
		1 & - e^{i l}\\
		e^{-i l} & 1
	\end{pmatrix},
\end{align}
which results in
\begin{align}
	S=\frac{1}{2}\left(
	\begin{array}{cc}
		e^{i \alpha _1}+e^{i \alpha _2-2 i l} & e^{i \left(\alpha
			_1+l\right)}-e^{-i \left(l-\alpha _2\right)} \\
		e^{i \left(\alpha _1+l\right)}-e^{-i \left(l-\alpha _2\right)} &
		e^{i \alpha _2}+e^{i \left(\alpha _1+2 l\right)} \\
	\end{array}
	\right),\quad l=\frac{1}{2}\arccos\frac{E}{2t}.
\end{align}
This is the final form of S-matrix, which describes the magnetic breakdown near usual saddle point. By making the cut-off parameter $N_c$ large enough, we calculate the S-matrix with arbitrary precision.
The results of the calculation in comparison with exact expression as a function of energy are presented in Fig.2 of the main text. The figure demonstrates excellent agreement between exact analytic expression given by and numerical calculations with $N_c=2000$. 

\subsection{Monkey saddle point}
\label{sec:Monkey-saddle}
\begin{figure*}
	\centering
	\includegraphics[scale=0.5]{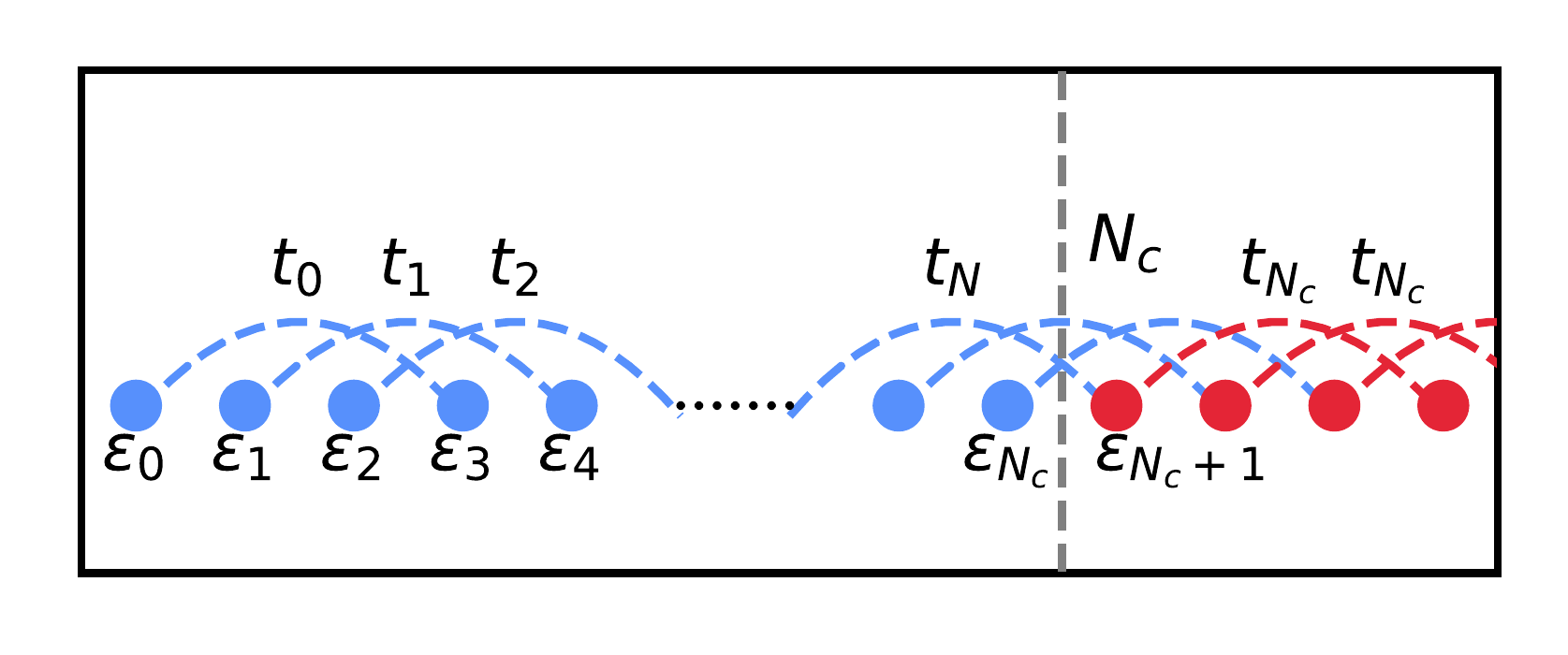}
	\caption{Effective 1D tight-binding model that represents the system of equations \eqref{eq:recursive-saddle-Monkey} for Monkey saddle with asymptotic truncation after Landau level index $n>N_c$.}
	\label{fig:Monkey-saddle-chain-model}
\end{figure*}

In this section, we extend the algorithm described above to the more complicated case of Monkey saddle point. The main complication arises due to the fact that the Hamiltonian is now third-order and we have to define $3\times3$ S-matrix between $3$ incoming and $3$ outgoing trajectories. Typically, the effective Hamiltonian in the vicinity of Monkey saddle point has the form \cite{Chamon2020,Yuan2020PRB}:
\begin{align}
	H_M=\left(k_x^3-3 k_x k_y^2\right),
\end{align}
where we omit constants for simplicity.
To obtain this operator in terms in ladder operators for a system under magnetic field, we perform symmetrization of the second term that makes the Hamiltonian Hermitian. The direct calculation shows that possible choices $\Pi_y \Pi_x \Pi_y$ and $\frac{1}{2}\left(\Pi_x \Pi_y^2+\Pi_y^2 \Pi_x\right)$ give the same result. In other words, we find in terms of ladder operators that the following expressions are identical:
\begin{align}\label{eq:pixpiypix}
	&- 2\sqrt{2} \Pi_y \Pi_x \Pi_y = \left(\hat{a}-\hat{a}^{\dagger}\right)\left(\hat{a}+\hat{a}^{\dagger}\right)\left(a-a^{\dagger}\right) 
	=\hat{a}^3 -\hat{a}^2 \hat{a}^{\dagger} +\hat{a}\hat{a}^{\dagger} \hat{a} -\hat{a}\hat{a}^{\dagger,2} -\hat{a}^{\dagger}\hat{a}^2+\hat{a}^{\dagger}\hat{a}\hat{a}^{\dagger}-\hat{a}^{\dagger,2} \hat{a}+\hat{a}^{\dagger,3} =\nn
	& (\text{we use commutation relation}\,\, [\hat{a},\hat{a}^{\dagger}]=1\,\,\text{and find}\,\, -\hat{a}^2 \hat{a}^{\dagger}-\hat{a}^{\dagger}\hat{a}^2=-2\hat{a} \hat{a}^{\dagger}\hat{a},\,\,-\hat{a}^{\dagger,2} \hat{a}-\hat{a}\hat{a}^{\dagger,2}=-2 \hat{a}^{\dagger}\hat{a}\hat{a}^{\dagger})\nn
	&=\hat{a}^3-\hat{a}\hat{a}^{\dagger}\hat{a}-\hat{a}^{\dagger}\hat{a}\hat{a}^{\dagger}+\hat{a}^{\dagger,3},\\
	& - \sqrt{2}\left(\Pi_x \Pi_y^2+\Pi_y^2 \Pi_x\right) = \hat{a}^3-\hat{a}\hat{a}^{\dagger}\hat{a}-\hat{a}^{\dagger}\hat{a}\hat{a}^{\dagger}+\hat{a}^{\dagger,3}.
\end{align}
However, the ambiguity of choice of symmetrization should be formally resolved to apply the procedure for more complicated saddle point Hamiltonians. We do this by starting from the tight-binding Hamiltonian, that has such saddle point, and introducing canonical momenta there. Using Eq.(18) from the main text, we find the following third-order term in expansion
\begin{align}
	H_{t r}(\mathbf{k})&=
	2 t\left(\sin \Pi_x a-\sin \frac{\Pi_x-\sqrt{3} \Pi_y}{2} a-\sin \frac{\Pi_x+\sqrt{3} \Pi_y}{2} a\right)\nn
	&\sim -\frac{t}{4}\left(\Pi_x^3-\Pi_x \Pi_y \Pi_y -\Pi_y \Pi_x \Pi_y -\Pi_y \Pi_y \Pi_x\right)a^3+O(a^5). 
\end{align}
Using above-written two relations, we simplified the symmetrized expression to Eq.(8) in the main text. However, in the general case one should keep full expression of symmetrized Hamiltonian for a particular lattice. Converting into matrix form, we find for Monkey saddle:
\begin{align}
	H_{M}&=-\frac{1}{2\sqrt{2}}[\left(\hat{a}+\hat{a}^{\dagger}\right)^3+3\left(\hat{a}-\hat{a}^{\dagger}\right)\left(\hat{a}+\hat{a}^{\dagger}\right)\left(a-a^{\dagger}\right)]\nn
	&=\frac{1}{2\sqrt{2}}\left(
	\begin{array}{cccccccc}
		0 & 0 & 0 & 4 \sqrt{6} & 0 & 0 & 0 & \dots \\
		0 & 0 & 0 & 0 & 8 \sqrt{6} & 0 & 0 & \dots \\
		0 & 0 & 0 & 0 & 0 & 8 \sqrt{15} & 0 & \dots \\
		4 \sqrt{6} & 0 & 0 & 0 & 0 & 0 & 8 \sqrt{30} & \dots\\
		0 & 8 \sqrt{6} & 0 & 0 & 0 & 0 & 0 & 4 \sqrt{n(n+1)(n+2)} \\
		0 & 0 & 8 \sqrt{15} & 0 & 0 & 0 & 0 & \dots \\
		0 & 0 & 0 & 8 \sqrt{30} & 0 & 0 & 0 & \dots \\
		\dots & \dots & \dots & \dots & 4 \sqrt{n(n+1)(n+2)} & \dots & \dots & \dots \\
	\end{array}
	\right).
\end{align}
Searching the solution in the form \eqref{eq:state-in-osc-basis} of decomposition of wave function in oscillator basis states, we find the following system of coupled equations:
\begin{align}\label{eq:recursive-saddle-Monkey}
	&2\sqrt{2}E\phi_0 - 4\sqrt{6}\phi_3=0,\nonumber\nn
	&2\sqrt{2}E\phi_1 - 8\sqrt{6}\phi_4=0,\nonumber\nn
	&2\sqrt{2}E\phi_2 - 8\sqrt{15}\phi_5=0,\nonumber \nn
	&2\sqrt{2}E\phi_3 - 4\sqrt{6}\phi_0 - 8\sqrt{30}\phi_6=0,\nonumber\nn 
	&\dots\nonumber\nn
	&2\sqrt{2}E \phi_n - 4\sqrt{n(n-1)(n-2)}\phi_{n-3} - 4\sqrt{(n+1)(n+2)(n+3)} \phi_{n+3}=0.
\end{align}
Following the procedure of mapping on 1D tight-binding model, we find that this system in turn converts into three decoupled chains with only nearest neighbor hoppings in each (see Fig.~\ref{fig:Monkey-saddle-chain-model} and Fig.~2 from the main text). 

Next steps require the introduction of the cut-off parameter $N_c$ and classification of the modes appearing for higher indices $n>N_c$. In the case of the model with three decoupled chains the Bloch eigenstates are
\begin{align}
	\Psi_j(l)=\frac{1}{M} \sum_{n=N_c+1}^{\infty} e^{-3 i l n}|3 n+i\rangle, \varepsilon_{j}(l)=-E+2 t_{N_c} \cos 3 l,\quad j=0,\,1,\,2.
\end{align}
In this case, the size of the unit cell is $3l$.
Acting with the creation operator $\hat{a}^{\dagger}$, we find the following expression for the matrix in the basis of these states:
\begin{align}
	a_{\Psi, i n}^{\dagger}=\left(\begin{array}{ccc}
		0 & 0 & e^{3 i l} \\
		1 & 0 & 0\\
		0 & 1 & 0
	\end{array}\right).
\end{align}
Comparing to the matrix representation of $a_{\Psi, i n}^{\dagger}$ for usual van Hove singularity \eqref{eq:shift-operator-log-vhs}, we see that it has similar structure with shift of nodes by one index in the unit cell until the period reached. The eigenvalues and eigenvectors of this matrix are
\begin{align}\label{eq:Monkey-saddle-eigenvectors}
	\lambda_{1,2,3}=e^{il},\,\,e^{il\pm 2\pi i/3},\quad \chi_1=\frac{1}{\sqrt{3}}\begin{pmatrix}
		1\\
		e^{-il}\\
		e^{-2il}
	\end{pmatrix},\,\chi_2=\frac{1}{\sqrt{3}}\begin{pmatrix}
		e^{i l + 2 \pi i / 3}\\
		1\\
		e^{-il-2\pi i/3}
	\end{pmatrix}\,\chi_3=\frac{1}{\sqrt{3}}\begin{pmatrix}
		e^{2 i l + 2 \pi i / 3}\\
		e^{il-2\pi i/3)}\\
		1
	\end{pmatrix}
\end{align}
In the limit of $E\ll 2 t_{N_c}$, we find that momenta of the modes are approximately at the effective Brillouin zone edge,
\begin{align}
	l=\frac{1}{3}\arccos\frac{E}{2 t_{N_c}}\approx \frac{\pi}{6},
\end{align}
and the phases of eigenvalues $\lambda_{1,2,3}$ correspond to the angles at which scattering states come from the quasiclassical regions for a Monkey saddle (see Fig.~2(a) in the main text).

The eigenvectors in Eq.~\eqref{eq:Monkey-saddle-eigenvectors} define the rotation to the basis with definite angles. Combining these eigenvectors into unitary transformation matrix, we then apply the basis transformation to the diagonal scattering matrix obtained from the numerical calculation,
\begin{align}
	S_{\Psi}=\text{diag}(e^{i\alpha_1},e^{i\alpha_2},e^{i\alpha_3}).
\end{align}
The diagonal structure of this matrix is a result of chain decoupling in the effective tight-binding model.
We present the result of numerical calculation in Fig.~2 of the main text.

\subsection{ High-order saddle points with different powers in effective dispersion: $A_3$ saddle point and regularization}
Finally, we discuss the case of $A_3$ saddle point, which has two trajectories coming close in the magnetic breakdown region, but the effective Hamiltonian has different powers in leading order for $k_x$ and $k_y$. \begin{figure*}
	\centering
	\includegraphics[scale=0.5]{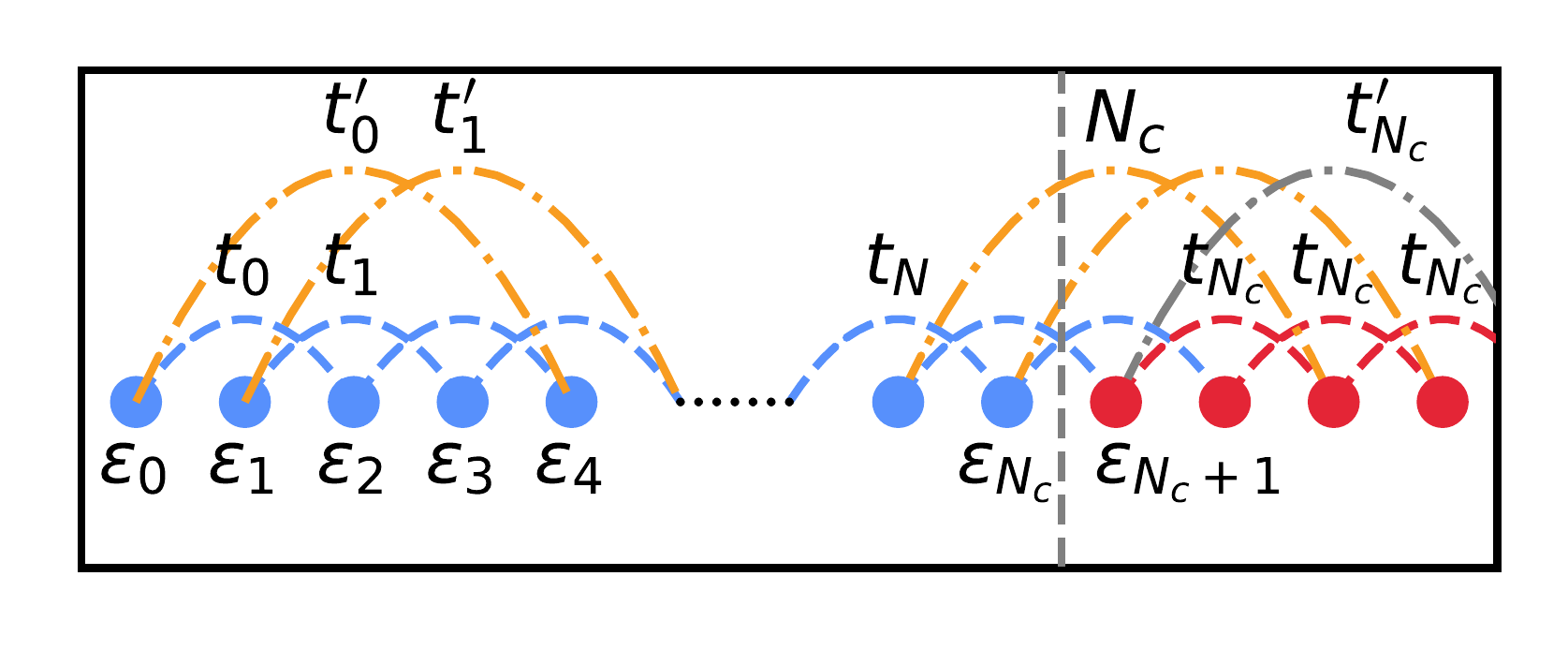}
	\caption{Effective 1D tight-binding model that represents the system of equations \eqref{eq:recursive-saddle-A3} for $A_3$ saddle with asymptotic truncation after Landau level index $n>N_c$. Note that now next nearest neighbor hoppings are added in each chain. }
	\label{fig:chain-model-A3-vHs}
\end{figure*}
The analysis of this point requires several additional steps, that can be used together with algorithm from previous section to analyze every other saddle point.
There are two complications that appear for such system comparing to usual saddle point discussed in Sec.~\ref{sec:supplement-log-vHs}. The effective Hamiltonian of the model is given by
\begin{align}
	H_{A_3}= \alpha k_x^2 a^2-k_y^4 a^4.
\end{align}
As we see, larger the deviation of momenta is from the center of saddle point, $k_x=k_y=0$, the closer trajectory angles are to zero. In other words, the constant energy curve solution $k_y(k_x,E)$ behaves asymptotically as $\sqrt{k_x}$ and never reaches the $k_x$-independent derivative. This introduces a problem of classification of modes by angle of incoming quasiclassical particle, the central ingredient used in previous cases. In fact, such problem may lead to unexpected numerical artifacts appearing in the final scattering matrix. We resolve this problem by introducing sub-leading terms into the Hamiltonian to make the highest polynomial powers for $k_x$ and $k_y$ identical. In our case this would lead to the following modification of original Hamiltonian:
\begin{align}\label{eq:A3-modified}
	\tilde{H}_{A_3}=\alpha k_x^2 a^2 +\beta k_x^4 a^4 -k_y^4 a^4,\quad \beta> 0.
\end{align}  
Here, the condition $\beta>0$ enforces the trajectories to have different asymptotic angles, that are found from the equation $\tan^4 \phi_k=\beta$, and are equal to $\pm \arctan \beta^{1/4},\,\,\pi\pm\arctan \beta^{1/4}$.
The ladder operator version of this Hamiltonian is 
\begin{align}
	\tilde{H}_{A_3}=\alpha\left(\hat{a}+\hat{a}^{\dagger}\right)^2+\beta\left(\hat{a}+\hat{a}^{\dagger}\right)^4-\left(\hat{a}-\hat{a}^{\dagger}\right)^4.
\end{align}
The main difference comparing to the previously discussed cases is that this Hamiltonian contains both second and fourth order terms, which implies the existence of two kinds of hopping terms. The system of recursive equations in this case is 
\begin{align}
	\label{eq:recursive-saddle-A3}
	&E\phi_0-(\phi _0 (\alpha+3 \beta-3)+\sqrt{2} \phi _2 (\alpha+6 \beta+6)+2 \sqrt{6} (\beta-1) \phi _4)=0,\nn
	&E \phi_1-(3 \phi _1 (\alpha+5 \beta-5)+\sqrt{6} \phi _3 (\alpha+10 \beta+10)+2 \sqrt{30} (\beta-1) \phi _5)=0,\nn
	&\dots\nn
	&E\phi_n-\left(\sqrt{(n-1) n} \phi _{n-2} (\alpha+2 (\beta+1) (2 n-1))+\phi _n (2 \alpha n+\alpha+3 (\beta-1) (2 n (n+1)+1))+\right.\nn
	&\left.+\sqrt{(n+1) (n+2)} \phi _{n+2} (\beta+2 (\beta+1) (2 n+3))+(\beta-1)
	\sqrt{\prod_{j=0}^{3}(n-j)} \phi _{n-4}+(\beta-1) \sqrt{\prod_{j=1}^{4}(n+j)} \phi _{n+4}\right)=0
\end{align}
After mapping we obtain the system with two decoupled chains, see Fig.~\ref{fig:chain-model-A3-vHs}.
Now each of these chains contains both nearest neighbor $t_i$ and next nearest neighbor $t_i^{\prime}$ hopping terms as well as index-dependent on-site terms $\epsilon_n$. In the case of more general saddle points one might obtain even more far-distanced next-next-...-nearest hoppings according to polynomial powers in the Hamiltonian. The basis of plane waves in the truncated region is built in the same way as before, thus rotation to the basis with proper angles is done via the same procedure of diagonalization of ladder operator $\hat{a}^{\dagger}$ acting on propagating modes. However, the dispersion relation of plane wave modes is more complicated, in this particular case it has the form:
\begin{align}
	\epsilon_{1,2}(l)=-E+E_{N_c}+2t_{N_c}\cos2l+2t_{N_c}^{\prime}\cos 4l.
\end{align} 
Expressing momentum from this equation for $\epsilon(l)=0$ condition, we obtain proper angles of the modes from $a_{\Psi}^{\dagger}$ operator given by Eq.~\eqref{eq:ladder-operator-vhs-eigenvectors}. 

\begin{figure*}
	\centering
	\includegraphics[scale=0.5]{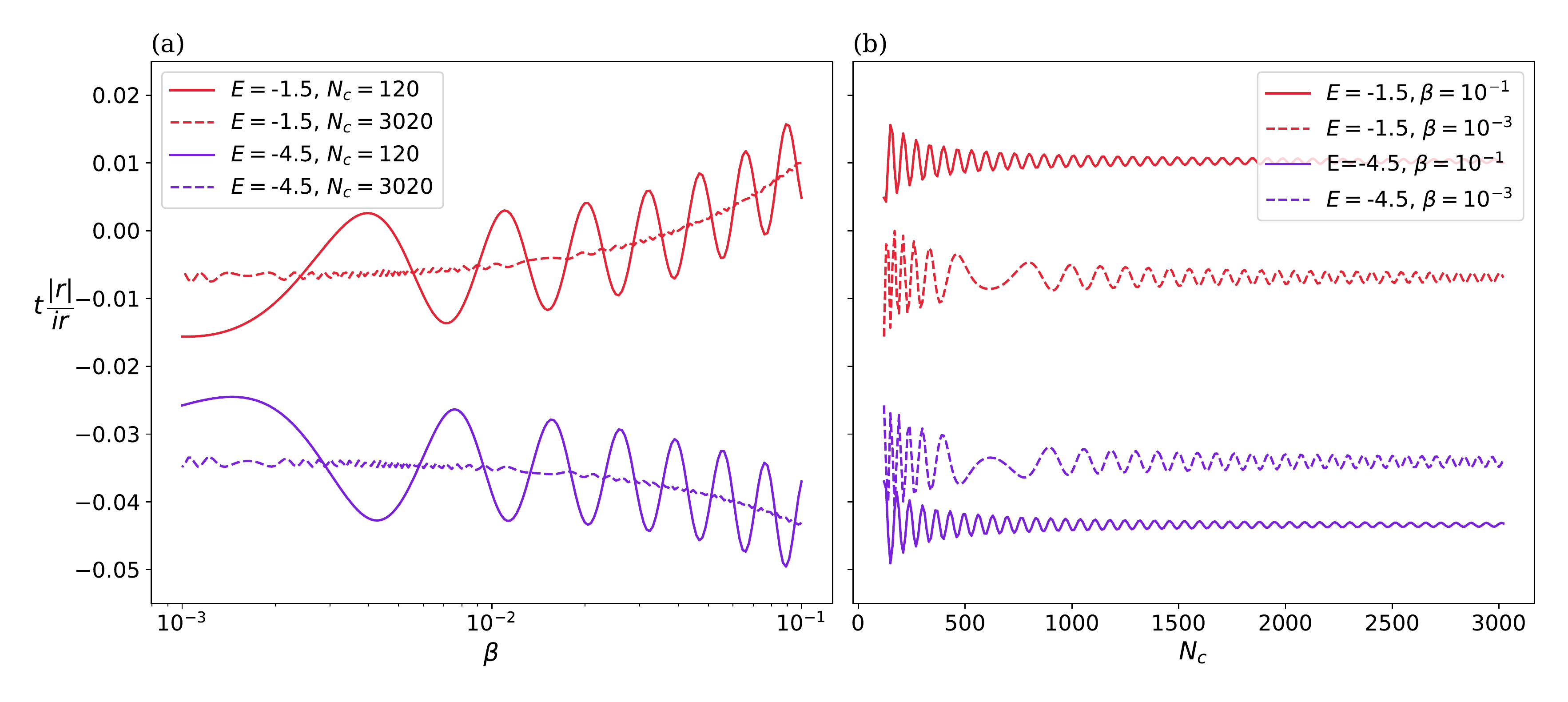}
	\caption{Convergence of absolute values of S-matrix elements ($t$-transmission, $r$-reflection) for $A_3$ saddle point with different values of regularization parameter $\beta$ in Eq.~\eqref{eq:A3-modified}. The coefficients $a=1$ and $\alpha=1$ are taken for simplicity. Two panels demonstrate convergence with different (a) $N_c$ cutoff parameters as function of $\beta$, (b) $\beta$ regularizing coefficients as function of $N_c$.}
	\label{fig:A3-convergence-B-comparison}
\end{figure*}

As a result of the regularization procedure, we have to introduce an additional parameter $\beta$ into the lowest-order effective Hamiltonian. We check this by taking this parameter small enough the convergence of results is reachable. We compare the results for absolute value of the scattering matrix elements fixing $A=1$ and taking different $\beta\ll 1$ in Eq.~\eqref{eq:A3-modified} The values of scattering matrix elements are shown in Fig.~\ref{fig:A3-convergence-B-comparison}. For small enough $\beta$ the results converge rapidly. 

\section{Magnetic translation operators and spectral equations for coherent networks \label{app:on}}
In this section of Supplemental material, we present the main technical steps that are required to describe the problem of two-dimensional Bloch electrons in presence of magnetic field applied perpendicularly to the system at energies in the vicinity of van Hove singularity, when a coherent orbit network is formed. We demonstrate our approach in cases of square and triangular lattices that exhibit three different types of van Hove singularities. Before proceeding to the derivation of spectral equations for particular systems, we briefly recall the translation symmetry group in the presence of magnetic field. 

\subsection{Magnetic translation group \label{app:magtrans}}
The general aspects of the translation symmetry group theory in the presence of magnetic field were discussed in Ref.~\cite{Fischbeck1970}. Here, we recall the key definitions and properties of Bloch electrons in the presence of magnetic field. 

The key operator that commutes with a Hamiltonian and therefore identifies the eigenstates and corresponding eigenvalues is the magnetic translation operator
\begin{align}
	\hat{T}_{\vec{R}}^{M}=\exp\{\frac{i}{\hbar} \left(\vec{p}+\frac{e}{c}\overline{\vec{A}}(\vec{R})\right)\vec{R}\},\quad \overline{\vec{A}}(\vec{R})=\vec{A}(\vec{R})+\vec{R}\times\vec{B},
\end{align}
with $\vec{p}$ being a momenta operator.
To find an analytical solution of the problem, commensurability between lattice and magnetic translation periods is required. This condition, expressed in terms of the magnetic translation operators, reads $\hat T^{M}_{N\vec{R}}\Psi=\Psi$, where $N$ is large integer and $\vec{R}$ an arbitrary lattice vector. The periodic condition implies $[\hat T^{M}_{NR},\hat T^{M}_{R'}]=0$ and can be equally written as a rationality condition on magnetic field flux through the elementary unit cell
\begin{align}
	\Phi = B |\vec{R}_1 \times \vec{R}_2| = B\frac{(2\pi)^2}{|\vec{b}_1 \times \vec{b}_2|}=\frac{q}{p} \frac{2\pi \hbar c}{e}=\frac{q}{p}\Phi_0,
\end{align}
where numbers $q,p$ are co-prime integers, $p$ is divisor of $N$, and $\Phi_0=\frac{2\pi e}{c}$ is a flux quantum. Later on we shall concentrate on the case of $q=1$ for which magnetic field values form a dense set for large $p$ and small magnetic fields. Other values of $q$ can be analyzed in a similar manner.

For the set of magnetic translation operator to form a group, we need to adjust its definition by adding a phase factor \cite{Fischbeck1970}
\begin{align}
	\hat T^{M}_{m, \vec{R}} = e^{i\pi (2m - jk)/p}\hat{T}_{\vec{R}}^{M}; \quad \vec{R} = j\vec{R}_1 + k\vec{R}_2,
\end{align}
where $R_{1}$, $R_{2}$ are elementary unit vectors. This group has $N^2$ $p$-dimensional irreducible representations, classified by quasimomenta eigenvalues $\vec{q} = \frac{1}{N}(n_1 \vec{b}_1 + n_2 \vec{b}_2), n_1,n_2 = 0,...,\frac{N}{p}-1$ and corresponding eigenstates $\Psi_{\alpha\vec{q}s}$ as follows  \cite{Fischbeck1970}
\begin{align}
	H \Psi_{\alpha\vec{q}s} = \mathcal{E}_\alpha(\vec{q})\Psi_{\alpha\vec{q}s}; \quad \hat T^{M}_{m, \vec{R}} \Psi_{\alpha\vec{q}s} = e^{i(\vec{q} + \frac{s}{p}\vec{b}_1)\vec{R}}\Psi_{\alpha\vec{q}s};  \quad s=0,...,p-1 
\end{align}
This relation defines the translational symmetry of the problem and may be considered as an analogy of Bloch's theorem.
\subsection{The wave function dependence on eigenvalues of the magnetic translation operator \label{app:quasimom}}
As described in the main text a general structure of the wave function expressed in the gauge invariant space reads \cite{Zilberman1957,Fischbeck1970,Alexandradinata2018} 
\begin{align}
	\Psi(k_x)&=\sum_{l, k, j}^{\infty}  \alpha_j e^{i(p_l l+ p_kk)} e^{i l_B^2 [k_x l b_{2,y}  -\frac{l^2}{2}b_{2,x}b_{2,y}]} \Psi_{ZF}^j{(k_x-kb_{1,x}-lb_{2,x})},
\end{align}
where indices $(l,k)$ enumerate cells in the extended Brillouin zone scheme, $j$ identifies different branches of the wave function in the single cell, and $\Psi_{ZF}^{E,j}(k_x)$ stands for ZF function, defined in the $(0,0)$-cell.

Phase factor $e^{i(p_l l+ p_kk)}$ defines the translational symmetry of the system and therefore must be expressed in terms of quasimomenta eigenvalues of magnetic translational operator $\vec{q}$. To find the connection between $p_l,p_k$ and $\vec{q}$ we project the solution on the representation space of magnetic translation group using a projection operator $P^{\vec{q}}$ \cite{Fischbeck1970}: 
\begin{align}
	\Psi_{\mathbf{q}}(k_x)=P^{\mathbf{q}} \Psi(k_x),\quad P^{\mathbf{q}}=C_{\vec{q}}\sum\limits_{k',l'=-\infty}^{\infty}e^{-i\vec{q} \cdot (\vec{b}\times \vec{ \lambda}) } e^{i\pi p k'l'+i (\vec{k}-\frac{e}{c}\vec{A}) \cdot (\vec{b}\times \vec{\lambda})},\quad  \vec{b} = k' \vec{b}_1 + l' \vec{b}_2
\end{align}
where $C_{\vec{q}}$ is a normalization constant and $\boldsymbol{\lambda}=(0,0,l_B^2)$ is a vector oriented along magnetic field . It is convenient to choose a system of coordinates where $\vec{b}_1 = (b_{1,x}, 0,0)$ and $\vec{b}_2 = (b_{2,x}, b_{2,y},0)$ where the projection operator expressed in the gauge invariant space $k_x = (\vec{p} + \frac{e}{c}\vec{A})_x, -\frac{i}{l_B^2}\partial_{k_x} = (\vec{p}+ \frac{e}{c}\vec{A})_y$ reads
\begin{align}
	P^{\mathbf{q}}=C_{\vec{q}}\sum\limits_{k',l'}e^{-i l_B^2  l' (q_1b_{2,y}-q_2b_{2,x})}e^{i l_B^2  l' (b_{2,y}k_x+i\frac{b_{2,x} }{l_B^2 }\partial_{k_x})}e^{i l_B^2  k'  q_2b_{1,x}}e^{-k'b_{1,x}\partial_{k_x}}
\end{align}
First, we calculate the projection part depending on $k '$
\begin{align}
	&e^{i l_B^2  k'  q_2b_{1,x}}e^{-k'b_{1,x}\partial_{k_x}}\sum_{l, k, j}^{\infty} \alpha_j e^{i(p_l l+ p_kk)} e^{i l_B^2 [k_x l b_{2,y}  -\frac{l^2}{2}b_{2,x}b_{2,y}]} \Psi_{ZF}^{j}{(k_x-kb_{1,x}-lb_{2,x})}\nonumber\\
	&= e^{i l_B^2  k'  q_2b_{1,x}}\sum_{l, k, j}^{\infty} \alpha_j e^{i(p_l l+ p_kk)} e^{i l_B^2 [(k_x-k' b_{1,x}) l b_{2,y}  -\frac{l^2}{2}b_{2,x}b_{2,y}]} \Psi_{ZF}^{j}{(k_x-(k+k')b_{1,x}-lb_{2,x})}\nonumber\\
	&= e^{i   k'  [l_B^2q_2b_{1,x}-p_k]}\sum_{l, k, j}^{\infty} \alpha_j e^{i(p_l l+ p_k k)} e^{i l_B^2 [k_x l b_{2,y}  -\frac{l^2}{2}b_{2,x}b_{2,y}]} \Psi_{ZF}^{j}{(k_x-kb_{1,x}-lb_{2,x})},
\end{align}
where in the second line we used $e^{ a\partial_{k_x}}f(k_x) = f(k_x+a)$ and in the last line we renamed $k \to k+k'$ and used $e^{il_B^2 b_{1,x}b_{2,y}}=e^{il_B^2 |\vec{b}_1 \times \vec{b}_2|}=e^{i2\pi p}=1$. Now, we calculate the second part of projection depending on $l'$
\begin{align}
	&e^{-i l_B^2  l' (q_1b_{2,y}-q_2b_{2,x})}e^{i l_B^2  l' (b_{2,y}k_x+i\frac{b_{2,x} }{l_B^2 }\partial_{k_x})}\sum_{l, k, j}^{\infty} \alpha_j e^{i(p_l l+ p_kk)} e^{i l_B^2 [k_x l b_{2,y}  -\frac{l^2}{2}b_{2,x}b_{2,y}]} \Psi^{j}_{ZF}{(k_x-kb_{1,x}-lb_{2,x})}\nonumber\\
	&=e^{-i l_B^2 ( l' (q_1b_{2,y}-q_2b_{2,x})+  \frac{l'^2}{2} b_{2,x}b_{2,y}-  l' b_{2,y}k_x)}e^{-l'b_{2,x}\partial_{k_x}}\sum_{l, k, j}^{\infty} \alpha_j e^{i(p_l l+ p_kk)} e^{i l_B^2 [k_x l b_{2,y}  -\frac{l^2}{2}b_{2,x}b_{2,y}]} \Psi^{j}_{ZF}{(k_x-kb_{1,x}-lb_{2,x})}\nonumber\\
	&=e^{i l_B^2  (l' b_{2,y}k_x-  l' (q_1b_{2,y}-q_2b_{2,x})-  \frac{l'^2}{2} b_{2,x}b_{2,y})}\sum_{l, k, j}^{\infty} \alpha_j e^{i(p_l l+ p_kk)} e^{i l_B^2 [(k_x-l'b_{2,x})l b_{2,y}  -\frac{l^2}{2}b_{2,x}b_{2,y}]} \Psi_{ZF}^{j}{(k_x-kb_{1,x}-(l+l')b_{2,x})}\nonumber\\
	&= e^{-i l'[l_B^2   (q_1b_{2,y}-q_2b_{2,x})-p_l]}\sum_{l, k, j}^{\infty} \alpha_j e^{i(p_l l+ p_kk)} e^{i l_B^2 [k_xl b_{2,y}  -\frac{l^2}{2}b_{2,x}b_{2,y}]} \Psi_{ZF}^{j}{(k_x-kb_{1,x}-lb_{2,x})}
\end{align}
Finally we see that the projecting results in
\begin{align}
	\Psi_{\mathbf{q}}(k_x)=C_{\vec{q}}\sum\limits_{k',l'=-\infty}^{\infty}e^{-i l' [l_B^2 (q_1b_{2,y}-q_2b_{2,x})+p_l]} e^{i   k'  [l_B^2q_2b_{1,x}-p_k]} \Psi(k_x).
\end{align}
The summation over $k',l'$ establishes the connection between $p_k, p_l$ and $\vec{q}$ as $p_k = l_B^2q_2b_{1,x}$ and $p_l = l_B^2(q_2b_{2,x}-q_1b_{2,y})$. 
Utilizing this relation, we find a general form of the wave function as a eigenfunction of the magnetic translation operator
\begin{align}
	\Psi_{\vec{q}}{(k_x)}&=\sum_{l, k, j}^{\infty}  \alpha_j e^{i l_B^2 ([q_2b_{2,x}-q_1b_{2,y}]l+ q_2b_{1,x}k)} e^{i l_B^2 [k_x l b_{2,y}  -\frac{l^2}{2}b_{2,x}b_{2,y}]} \Psi_{ZF}^{j}{(k_x-kb_{1,x}-lb_{2,x})}
\end{align}
To define $\alpha_i$ coefficients and allowed energy values, we need to connect pieces of the wave function in neighbouring cells of the Brillouin zone as well as pieces corresponding to the different parts of the orbit in the single cell. In the following sections, we perform this for the cases of square and triangular lattices by solving a scattering problem near the van Hove singularities.

\subsection{Square lattice: derivation of spectral equation for orbit network}
In this section, we present the technical details of derivation of the orbit network spectral equation for the simplest case of square lattice. The Brillouin zone of the square lattice contains saddle points at the X-points. The tight-binding Hamiltonian in the case where only nearest neighbor hopping is taken into account reads
\begin{align}\label{eq:H-square-usual}
	H_{sq}=-2 t (\cos(k_x a)+\cos(k_y a)),
\end{align}
with $a$ being a lattice constant and $t$ is hopping parameter. This Hamiltonian exhibits the saddle points at $X$-points in Brillouin zone
\begin{figure*}
	\centering
	$\begin{array}{cc}
		& \includegraphics[scale=0.45]{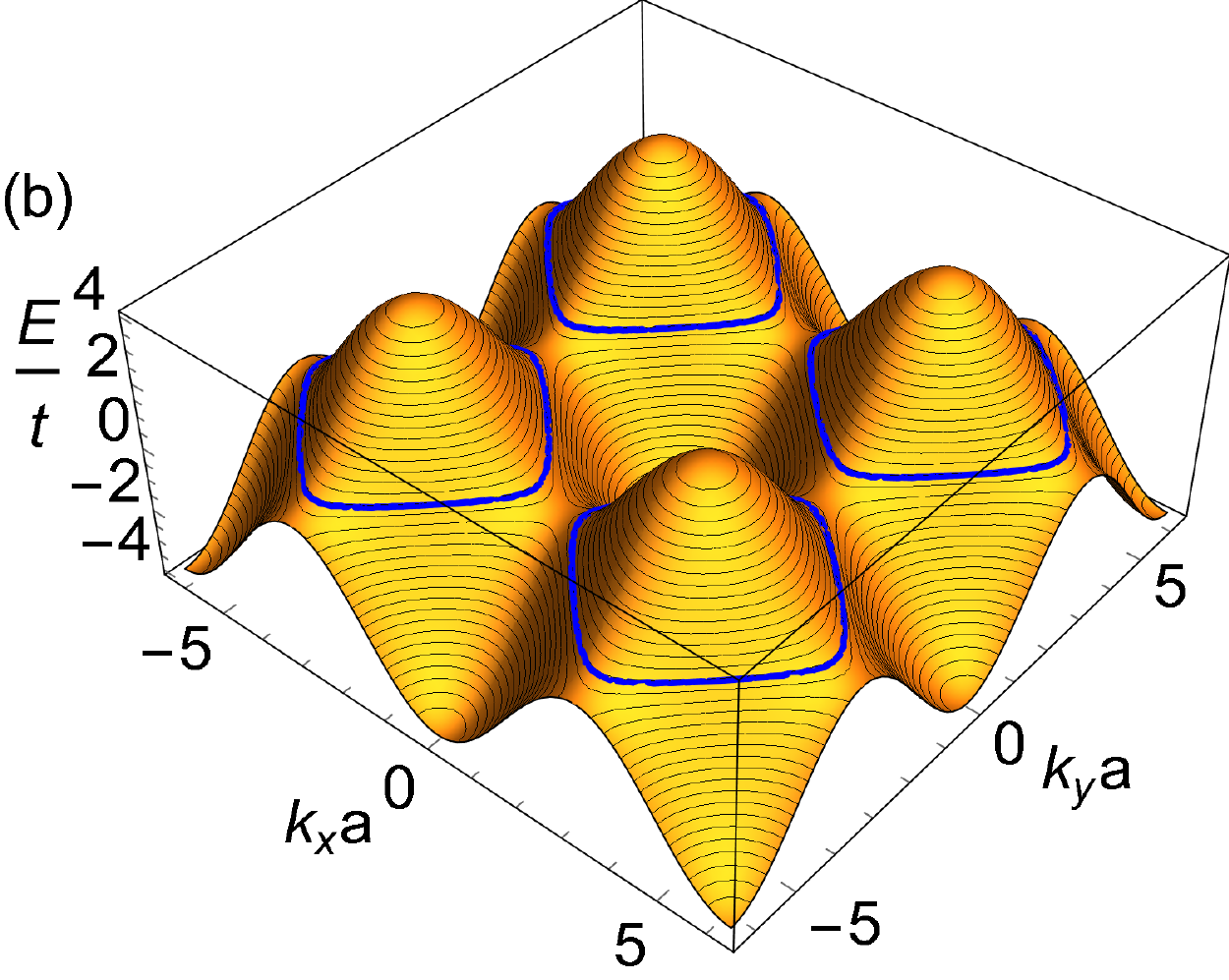}\\
		\smash{\raisebox{.5\normalbaselineskip}{\includegraphics[scale=0.14]{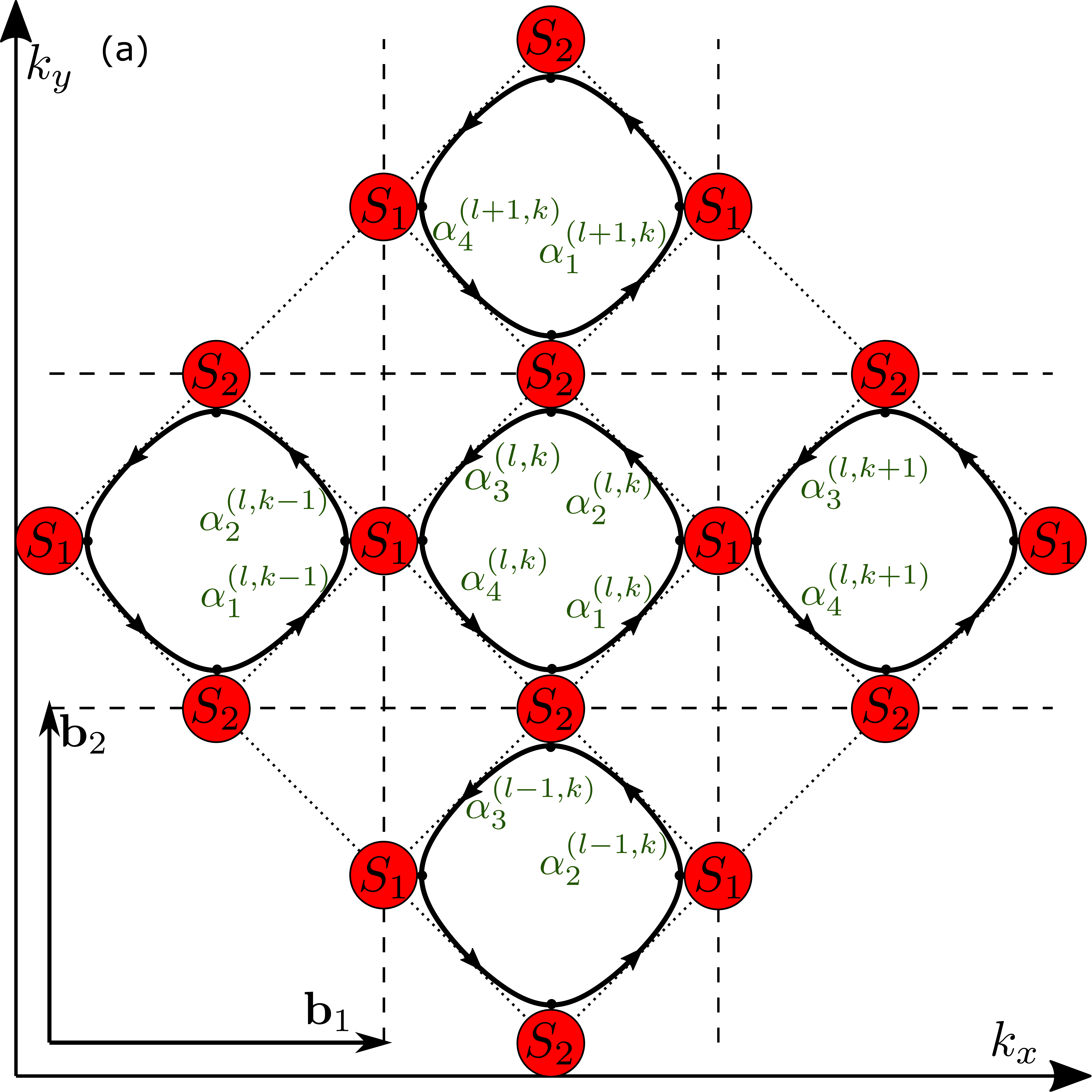}}} & \includegraphics[scale=0.45]{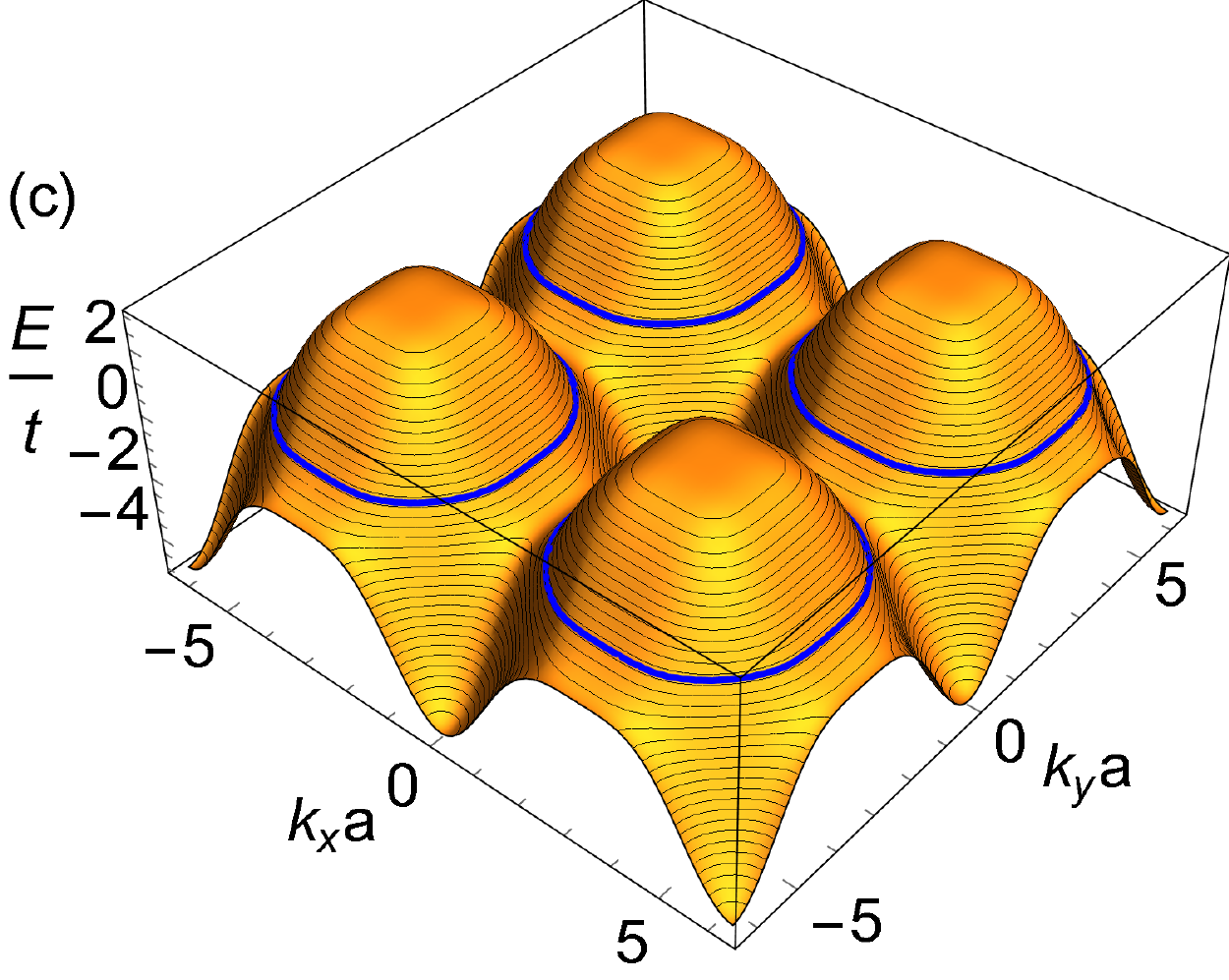}
	\end{array}$
	\caption{Panel (a): schematic geometry of the orbit network from Eq.\ref{eq:s2-wf-square} in the square lattice. Panel (b): tight binding energy dispersion of the square lattice with only nearest neighbor hoppings $t$ involved; constant energy contours and schematicly represented orbits at a particular fixed energy (blue lines) are highlighted. Panel (c): spectrum of the same square lattice as in (b) but with third nearest neighbor hoppings $t_3=1/4 t$ taken into account.}
	\label{fig:orbit-network-square-lattice}
\end{figure*}
\begin{align}\label{eq:series-decomp-square-X-usual}
	E_{\pi / a, 0}(\vec{k}) \approx-a^2 t [(k_x-\pi/a)^2-k_y^2],\quad E_{0, \pi / a}(\vec{k}) \approx a^2 t [k_x^2-(k_y-\pi/a)^2].
\end{align}
They result in usual logarithmic van Hove singularities in the density of states. One can modify Eq.~\eqref{eq:H-square-usual} by taking hopping $t_3$ to the third nearest neighbor into account 
\begin{align}\label{eq:H-square-modified}
	H_{sq, 2}=-2 t\left(\cos k_x a+\cos k_y a\right)-2 t_3\left(\cos 2 k_x a+\cos 2 k_y a\right),
\end{align}
and obtain high-order saddle points of cusp type $A_3$ \cite{Chamon2020} at each X-point for $t_3=1 / 4 t$:
\begin{align}\label{eq:series-decomp-square-X-A3}
	&E_{\pi / a, 0} \approx-\frac{1}{4} a^4 t q_x^4-\frac{5}{12} a^4 t q_y^4+2 a^2 t q_y^2-t, \nn
	&E_{0, \pi / a} \approx-\frac{5}{12} a^4 t q_x^4+2 a^2 t q_x^2-\frac{1}{4} a^4 t q_y^4-t.
\end{align}
Such saddle points result in high-order van Hove singularities with divergence exponent $1/4$. Recently, a slightly different version of this model with nearest-neighbor (NN), next NN and third NN hoppings was analyzed in Ref.~\cite{Han2023arxiv} with relation to enhanced nematicity effects. 

While we keep the hopping parameters isotropic (equal along x- and y-directions), the qualitative geometry of the coherent orbit network does not change due to saddle point type. It is schematically shown in Fig.~\ref{fig:orbit-network-square-lattice} together with tight-binding dispersion plots with highlighted equi-energy contours. 

To derive a solution for the orbit network, we note that each unit cell of the network in Fig.~\ref{fig:orbit-network-square-lattice} can be separated into four quasiclassical regions which can be described by WKB-type Zilberman-Fischbeck (ZF) wave functions, and four scattering regions of magnetic breakdown with corresponding scattering matrices $S_{1,2}$.

\begin{align}
	&\Psi_{\vec{q}}(k_x) = \sum\limits^{\infty}_{
		l,k=-\infty}
	e^{i (p_kk+p_ll)}e^{i l_B^2 l b_{2,y}k_x}\nn
	&\times\begin{cases}
		\alpha_{1}h_{1}(k_x)e^{il_B^2 \int\limits_{kb_{1,x}}^{k_x}d k_x k_y^{E,b}(k_x)}+\alpha_{2}h_{2}(k_x)e^{i l_B^2  \int\limits_{kb_{1,x}}^{k_x}d k_x k_y^{E,t}(k_x)},& kb_{1,x} < k_x <(k+\frac{1}{2}) b_{1,x}\\
		\alpha_{3}h_{3}(k_x)e^{il_B^2 \int\limits_{kb_{1,x}}^{k_y} d k_x k_y^{E,t}(k_x)}+\alpha_{4}h_{4}(k_x) e^{i l_B^2 \int\limits_{kb_{1,x}}^{k_x}d k_x k_y^{E,b}(k_x)},& (k-\frac{1}{2})b_{1,x} < k_x <k b_{1,x}
	\end{cases} \label{eq:s2-wf-square},
\end{align}
where $h_i(k_x)=\left(|\frac{\partial E(\vec{k})}{\partial k_y}|\right)^{-\frac{1}{2}}$. We choose zero of $k_x$ as point in the middle of single Brillouin zone. In each cell of reciprocal lattice, we have two solutions for $k^{E}_y(k_x)$ at constant energy which we call "top" $k_y^{E,t}(k_x)$ and "bottom" $k_y^{E,b}(k_x)$. The functional dependence $k_y^{E,t(b)}(k_x)$ is found directly from exact dispersion relation at a given energy. For example, in the case of Eq.~\eqref{eq:H-square-usual}, we find
\begin{align}
	k_{y}^{E,t(b)}(k_x,E)=\pm\frac{1}{a}\arccos\left[-\frac{E+2t\cos(k_y a)}{2t}\right],
\end{align}
while in most other cases the numerical evaluation has to be used.

To build the closed system of equations for coefficients $\alpha_j$, we use the magnetic breakdown S-matrix that couples two neighboring cells of the network. For the case of isotropic square lattice all S-matrices are identical, so the only difference comes from the geometric arrangement of cyclotron orbits coming to each magnetic breakdown region. We thus write the corresponding equations as `horizontal' scattering with $S_1$ matrix and `vertical' scattering with $S_2$ referring to the notation from Fig.~\ref{fig:orbit-network-square-lattice}(a). For the `horizontal' MB region we obtain:
\begin{align}\label{eq:Shorizontal-relation}
	\begin{pmatrix}
		\alpha_{2}e^{i\int \limits_{kb_{1,x}}^{(k+\frac{1}{2})b_{1,x}}d k_x k_y^{E,t}(k_x)}\\
		\alpha_{4}e^{ip_k-i\int \limits_{kb_{1,x}}^{(k+\frac{1}{2})b_{1,x}}d k_x k_y^{E,b}(k_x)}
	\end{pmatrix}=S_1\begin{pmatrix}
		\alpha_{1}e^{i\int \limits_{kb_{1,x}}^{(k+\frac{1}{2})b_{1,x}}d k_x k_y^{E,b}(k_x)}\\
		\alpha_{3}e^{ip_k-i\int \limits_{kb_{1,x}}^{(k+\frac{1}{2})b_{1,x}}d k_x k_y^{E,t}(k_x)}
	\end{pmatrix},
\end{align}
where we used that $h_{i}(k_x)$ are equal for all $i$ at scattering points.  Similarly for the `vertical' MB region we write:
\begin{align}\label{eq:Svertical-relation}
	\begin{pmatrix}
		\alpha_{1}e^{ip_l}\\
		\alpha_{3}
	\end{pmatrix}=S_2\begin{pmatrix}
		\alpha_{2}\\
		\alpha_{4}e^{ip_l}
	\end{pmatrix}.
\end{align}
Note that here, we used the relation $l_B^{2} b_{1,x}b_{2,y}=2\pi p$.

Now, taking into account the series decomposition of dispersion around saddle point in the case of isotropic square lattice (e.g. Eq.~\eqref{eq:series-decomp-square-X-usual} or Eq.~\eqref{eq:series-decomp-square-X-A3}) and the geometry of the orbits in Fig.~\ref{fig:orbit-network-square-lattice}, we obtain the rotation rules for the S-matrix ($\mathcal{R}$ and $\mathcal{T}$ are absolute values of reflection and transmission coefficients along the $k_x$ direction for $S_1$):
\begin{align}
	S_{1}=\begin{pmatrix}
		- i \mathcal{R} e^{i\varphi_r} & \mathcal{T} e^{i\varphi_t}\\
		\mathcal{T} e^{i\varphi_t} & - i \mathcal{R} e^{i\varphi_r}
	\end{pmatrix},\quad S_2=\begin{pmatrix}
		i \mathcal{T} e^{i\varphi_t} & \mathcal{R} e^{i\varphi_r}\\
		\mathcal{R} e^{i\varphi_r} & i \mathcal{T} e^{i\varphi_t}
	\end{pmatrix}.
\end{align}
These relations are obtained by noting that absolute values of transitions between part of trajectories $1\to 2$ and $3\to 4$ at point $S_1$ should be equal to $2\to 3$ and $4\to 1$ at point $S_2$.  The unitarity of S-matrix implies the following restrictions on S-matrix:
\begin{align}
	\mathcal{R}^2+\mathcal{T}^2=1,\quad i \mathcal{RT} e^{-i(\varphi_r-\varphi_t)}- i \mathcal{RT} e^{-i (\varphi_t-\varphi_r)}= 2 \mathcal{RT} \sin(\varphi_r-\varphi_t)=0.
\end{align}
The second relation implies that either $\varphi_r=\varphi_t+0,\,\pi,\,\dots$ or one of $r$ or $t$ is equal to zero. Indeed, we find that both cases are realized for the $A_3$ saddle point, which is an example of perfect transmission along the semiclassical orbit. We note that the relation $\det(S)=e^{i \varphi_{sc}}=e^{2i\varphi_r}=e^{2i\varphi_r}$ also follows from the unitarity.

Combining \eqref{eq:Shorizontal-relation} and \eqref{eq:Svertical-relation} into system of equations for $\alpha_{1,\dots 4}^{l,k}$ parameters, we find
\begin{align}
	\left(\begin{array}{cccc}
		- i \mathcal{R} e^{i\varphi_r} e^{i \Phi_b} & -e^{i \Phi_t} & \mathcal{T} e^{i\varphi_t} e^{i\left(p_k-\Phi_t\right)} & 0 \\
		\mathcal{T} e^{i\varphi_t} e^{i \Phi_b} & 0 & -i \mathcal{R} e^{i\varphi_r} e^{i\left(p_k-\Phi_t\right)} & -e^{i\left(p_k-\Phi_b\right)} \\
		-e^{i p_l} & i \mathcal{T} e^{i\varphi_t} & 0 & \mathcal{R} e^{i\varphi_r} e^{i p_l} \\
		0 & \mathcal{R} e^{i\varphi_r} & -1 & i \mathcal{T} e^{i\varphi_t} e^{i p_l}
	\end{array}\right) \boldsymbol{\alpha}=0,
\end{align}
where we used the short-hand notation
\begin{align}
	\Phi_{b(t)}=l_B^2\int_{k b_{1,x}}^{\left(k+\frac{1}{2}\right) b_{1,x}} d k_x k_y^{E,b(t)}\left(k_x\right).
\end{align}
The condition of existence of nontrivial solution gives the spectral equation:
\begin{align}
	([\mathcal{R} e^{i\varphi_r}]^2+[\mathcal{T} e^{i\varphi_t}]^2)^2 e^{i(\Phi_{b}-\Phi_{t})}+e^{-i(\Phi_{b}-\Phi_{t})}-\mathcal{R} e^{i\varphi_r}\mathcal{T} e^{i\varphi_t} \left(e^{ip_l}+e^{-ip_l}+e^{i(p_k-\Phi_{b}-\Phi_{t})}e^{-ip_l}+e^{-i(p_k-\Phi_{b}-\Phi_{t})}\right)=0
\end{align}
Using that $([\mathcal{R} e^{i\varphi_r}]^2+[\mathcal{T} e^{i\varphi_t}]^2)^2=\det S_1^2 = e^{2i\varphi_{sc}}$, $\Phi_b=l_B^2 b_{1,x} b_{2,y}-\Phi_t=2p\pi-\Phi_e$ (see Fig.~\ref{fig:orbit-network-square-lattice}) and the area inside the orbit is $l_B^2\mathcal{A}(E) =2(\Phi_t-\Phi_b)$, and obtain:
\begin{align}
	&\cos (\frac{l_B^2\mathcal{A}(E)}{2}-\varphi_{sc})=e^{i(\varphi_r-\varphi_t)}\mathcal{RT} \left(\cos
	\left(p_k\right)+\cos
	\left(p_l\right)\right)=\pm \mathcal{R}\mathcal{T} \left(\cos
	\left(p_k\right)+\cos
	\left(p_l\right)\right), 
\end{align}
where the sign depends on whether $\varphi_r=\varphi_t$ or $\varphi_r=\varphi_t+\pi$. Rewriting this equation in terms of magnetic translation operators eigenvalues, we find:
\begin{align}
	\cos\left(\frac{l_B^2\mathcal{A}(E)}{2}-\varphi_{sc}\right)=
	\pm \mathcal{RT} \left(\cos(l_B^2 q_1b_{2,y})+\cos(l_B^2 q_2b_{1,x})\right).
\end{align}
The cosine on the left defines the standard Lifshitz-Onsager quantization rule with slightly shifted Landau levels in the vicinity of the energy of van Hove singularity due to nonzero scattering phase. The right-hand side defines mini-band broadening and leads to oscillatory behavior.

\subsection{Triangular lattice with imaginary hoppings: orbit network connected via Monkey saddle points}
\begin{figure*}
	\centering
	\includegraphics[scale=0.08]{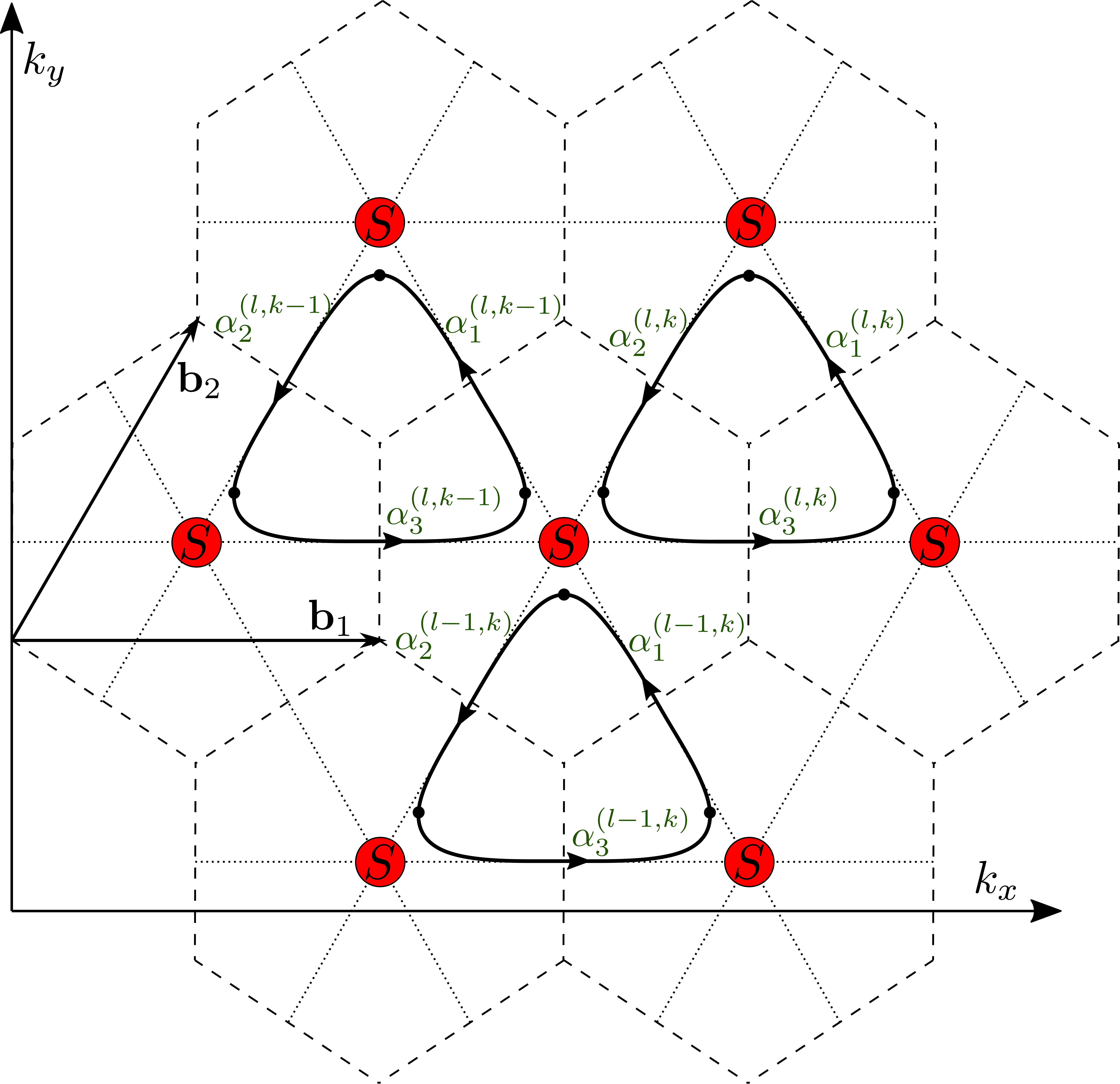}
	\caption{Schematic structure of the orbit network for triangular lattice with imaginary hoppings. The $\vec{b}_i$ vectors denote the basis in reciprocal space, red S circles label the positions of scattering regions with Monkey saddle. The black lines with arrows show example orbits with coefficients $\alpha_{i}^{l,k}$ that appear in decomposition of the wave function into Zilberman-Fischbeck wave functions between scattering regions. Indices $l$ and $k$ label the elementary unit cell in $k$-space with its position according to $\vec{b}_1$ and $\vec{b}_2$ vectors. Only the case of positive energies is shown as the negative energies can be obtain by using the symmetry of the model $E\to-E$ with $\vec{k}\to-\vec{k}$.}
	\label{fig:triangular-lattice-orbit-notation}
\end{figure*}

In this section, we analyze a case with a more complicated geometry of a triangular lattice with imaginary hopping parameters. The tight-binding Hamiltonian in the nearest neighbor approximation is given by Eq.~(18) in the main text.
Similar structure of dispersion with Monkey saddle might effectively appear in Moir\'{e} materials \cite{Hsu2021PRB}. We apply a $\pi/2$ clockwise rotation of coordinates $k_y \to k_x,  k_x \to -k_y$ to the system in order to have $\vec{b}_1 = (b_{1,x},0,0)$ as described in the previous sections. A schematic picture of the orbits that constitute a network is shown in Fig.~\ref{fig:triangular-lattice-orbit-notation}.
The general structure of a solution is described in the main text, in the particular case of triangular lattice the wave function has three different ZF-type terms in each unit cell and for $E>0$ reads:
\begin{align}\label{eq:Psi-trial-solution-triangular-lattice}
	\Psi_{\vec{q}}(k_x) = \sum\limits_{
		l,k}
	e^{i (p_kk+p_ll)}&e^{i l_B^2 [k_x l b_{2,y}  -\frac{l^2}{2}b_{2,x}b_{2,y}] } \Biggl[\alpha_{3}h_{3}(k_x)e^{i l_B^2 \int \limits_{k b_{1,x}+l b_{2,x}}^{k_x} d  k_x k_y^{E,3}(k_x)}, kb_{1,x}+lb_{2,x} < k_x < (k+1)b_{1,x}+lb_{2,x}\nonumber\\
	&+\begin{cases}
		\alpha_{2}h_{2}(k_x)e^{i l_B^2 \int \limits_{k b_{1,x}+(l+1)b_{2,x}}^{k_x} d  k_x k_y^{E,2}(k_x)}, &kb_{1,x}+lb_{2,x} < k_x < kb_{1,x}+(l+1)b_{2,x}\\
		\alpha_{1}h_{1}(k_x)e^{i l_B^2 \int \limits_{k b_{1,x}+(l+1)b_{2,x}}^{k_x} d  k_x k_y^{E,1}(k_x)},& kb_{1,x}+(l+1)b_{2,x} < k_x < (k+1)b_{1,x}+lb_{2,x}
	\end{cases}\Biggr].
\end{align}
The positions of scattering point are $k^{(l,k)}_{x,scatt}=kb_{1,x}+lb_{2,x}$.
The case of $E<0$ can be analyzed in the same way as $E>0$ by making use of a symmetry property of the dispersion (3) that stays the same for $E\to-E$ and $\vec{k}\to-\vec{k}$ replacement. Therefore, later on we concentrate on the case of $E>0$.

Now, we perform a derivation of spectral equation from the scattering equations that couple neighboring cells. The scattering equation at point $k^{(l,k)}_{x,scatt}$ has the form:
\begin{align}
	&\begin{pmatrix}
		\alpha_{1}e^{-ip_k}e^{i\Phi_1}\\
		\alpha_{2}e^{-ip_l}e^{-il_B^2\frac{b_{2,x}b_{2,y}}{2}}\\
		\alpha_{3}
	\end{pmatrix}=S\begin{pmatrix}
		\alpha_{1}e^{-ip_l}e^{-il_B^2\frac{b_{2,x}b_{2,y}}{2}}\\
		\alpha_{2}e^{-i \Phi_2}\\
		\alpha_{3}e^{-ip_k}e^{i \Phi_3},
	\end{pmatrix},\label{eq:scat}
\end{align}
where we used the relation $l_B^2 b_{1,x}b_{2,y} = 2\pi p$ and equality of $h_{i}(k_x)$ at the scattering point as well as the short-hand notation for $\Phi_i$ introduced in the previous section.

\begin{figure*}
	\centering
	\includegraphics[scale=0.085]{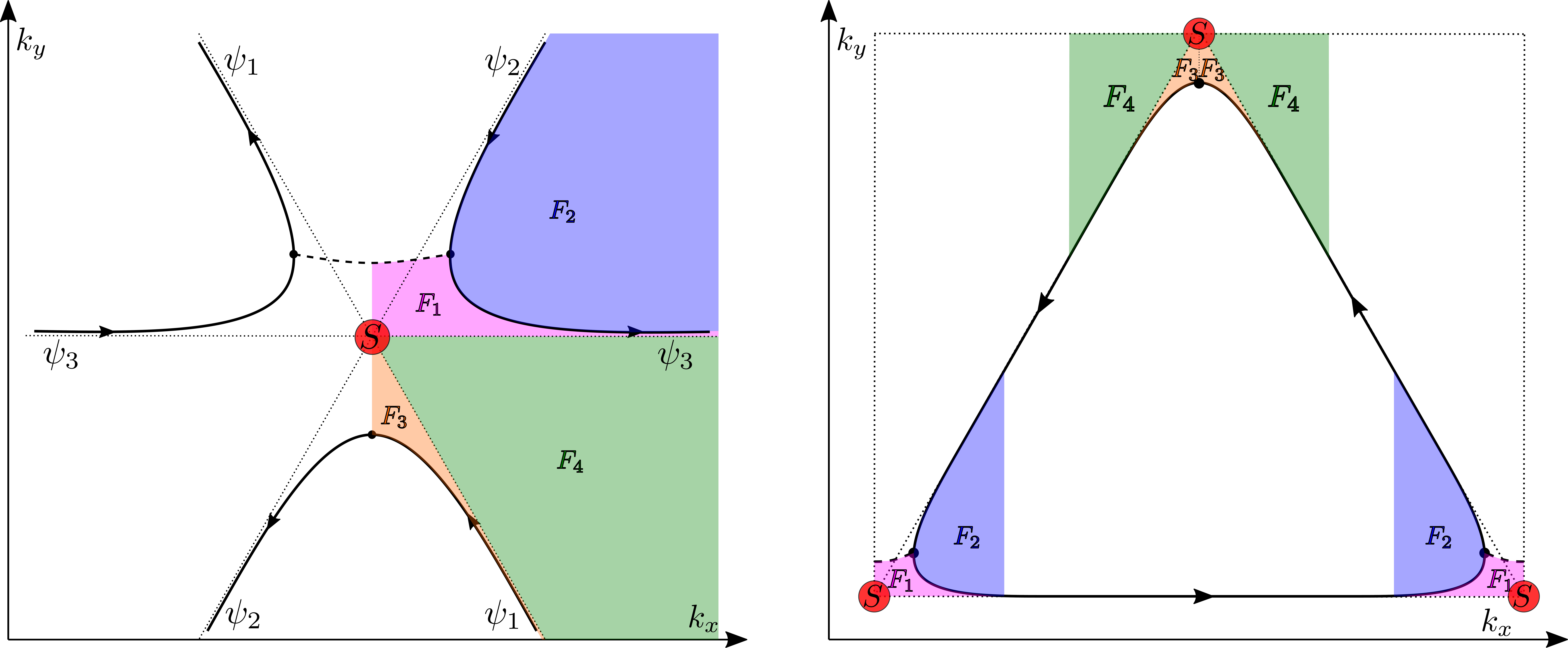}
	\caption{Definition of different phases according to the integration in momentum space for the scattering problem geometry around Monkey saddle point. Solid lines represent semiclassical trajectories at fixed energy near Monkey saddle, whereas dashed line depicts the real part of complex roots of Eq.~\eqref{eq:kx-roots} and dotted lines correspond to $E=0$ regime. The shaded areas are related to the constant and dynamical phases according to Eq.~\eqref{eq:F-definition-with-phases}.}
	\label{fig:Monkey-saddle-dynamical-phases}
\end{figure*}

Before we insert the numerically calculated S-matrix, there is one subtlety that should be taken into account: the scattering basis of incoming and outgoing modes in numerical approach \eqref{eq:Monkey-saddle-eigenvectors} is different from the one used in the orbit network \eqref{eq:Psi-trial-solution-triangular-lattice}. The difference between two pictures arises due to the fact that the numerical approach couples the incoming and outgoing modes in quasiclassical region at infinity $|\vec{k}|\to\infty$, while in the orbital network the modes are coupled at the scattering point itself  
\begin{align}
	\text{\small Orbit network:} \,\begin{pmatrix}
		\Psi^{out,1}_{ZF} (k=0)\\
		\Psi^{out,2}_{ZF} (k=0)\\
		\Psi^{out,3}_{ZF} (k=0)
	\end{pmatrix}=S\begin{pmatrix}
		\Psi^{in,1}_{ZF} (k=0)\\
		\Psi^{in,2}_{ZF} (k=0)\\
		\Psi^{in,3}_{ZF} (k=0)
	\end{pmatrix},\,\text{\small Numerical approach:} \, 	\begin{pmatrix}
		\Psi^{out,1}_{ZF} (k\to\infty)\\
		\Psi^{out,2}_{ZF} (k \to \infty)\\
		\Psi^{out,3}_{ZF} (k \to \infty)
	\end{pmatrix}^{\prime}=S'\begin{pmatrix}
		\Psi^{in,1}_{ZF} (k \to \infty)\\
		\Psi^{in,2}_{ZF} (k \to \infty)\\
		\Psi^{in,3}_{ZF} (k \to \infty)
	\end{pmatrix}^{\prime}.
\end{align}
Here, $\Psi^{i}_{ZF}$ and the primed $\Psi^{i,\prime}_{ZF}$ wave functions correspond to the different choices of normalization constant in the bases. Thus we continue with connecting the two types of scattering states in these problems by introducing dynamical phases as well as constant phase shifts that account for selected convention in basis definitions:
\begin{align}
	& \Psi_{ZF}^{in,i}\left(k_x\right)=e^{i \delta_i^{i n}} \Psi_{ZF}^{\prime i n,i}\left(k_x\right), \quad \Psi_{ZF}^{i n,i}\left(k_x \rightarrow \pm \infty\right)=e^{i \phi_{dyn}^{in,i}} \Psi_{ZF}^{in,i}(0) \\
	& \Psi_{ZF}^{out,i }\left(k_x\right)=e^{i \delta_i^{out}} \Psi_{ZF}^{'out ,i}\left(k_x\right), \quad \Psi_{ZF}^{ out ,i}\left(k_x \rightarrow \pm \infty\right)=e^{i \phi_{d y n}^{out,i}} \Psi_{ZF}^{in ,i}(0).
\end{align}
The phase factors $e^{i \delta^{i n}_{i}}, e^{i \delta^{out }_{i}}$ do not depend on energy and represent differences of the basis definitions, while $\phi_{d y n}^{i n, i}, \phi_{d y n}^{o u t, i}$ are dynamical phases that correspond to covered areas in momentum space of ZF wave functions $\psi_i^{i n}$ - see Fig.~\ref{fig:Monkey-saddle-dynamical-phases}. There, solid lines represent the real roots of equation 
\begin{align}\label{eq:kx-roots}
	k_y^3+3 k_y k_x^2=-E, 
\end{align}
defining semiclassical trajectory $k^E_y\left( k_x\right)$, while the dashed lines correspond to real parts of the complex roots in the intervals of $k_x$ with only one allowed semiclassical trajectory. The relation between phases and the shaded areas is the following:
\begin{equation}\label{eq:F-definition-with-phases}
	\left\{\begin{array}{l}
		\phi^{i n, 3}_{d y n}=-F_1 \\
		\phi^{i n,1}_{d y n}=-F_3-F_4 \\
		\phi^{i n, 2}_{d y n}=F_1+F_2
	\end{array}, \quad\left\{\begin{array}{l}
		\phi^{out,3 }_{d y n}=F_1 \\
		\phi^{out,1 }_{d yn}=-F_1-F_2 \\
		\phi^{o u t, 2}_{d y n}=F_3+F_4
	\end{array}\right.\right.
\end{equation}
By using the symmetry of Monkey saddle dispersion $E\to-E$ together with $\vec{k}\to-\vec{k}$, we notice that $\Psi_{ZF}^{out,i}\left(k_x, E\right)=\Psi_{ZF}^{in,i}\left(-k_x, -E\right)$ and $ \Psi_{ZF}^{\prime  out,i }\left(k_x, E\right)=\Psi_{ZF}^{in,i }\left(-k_x, -E\right)$, which gives $e^{i \delta_i^{i n}}=e^{i \delta_i^{ {out }}}=e^{i \delta_i}$.
Hence,
\begin{align}
	& S=\left(\begin{array}{ccc}
		e^{-i\left[\phi_{d y n, 1}^{ {out }}-\delta_1\right]} & 0 & 0 \\
		0 & e^{-i\left[\phi_{d y n, 2}^{o u t}-\delta_2\right]} & 0 \\
		0 & 0 & e^{-i\left[\phi_{d y n, 3}^{o u t}-\delta_3\right]}
	\end{array}\right) S^{\text {num }}\left(\begin{array}{ccc}
		e^{i\left[\phi_{d y n, 1}^{i n}-\delta_1\right]} & 0 & 0 \\
		0 & e^{i\left[\phi_{d y n, 2}^{i n}-\delta_2\right]} & 0 \\
		0 & 0 & e^{i\left[\phi_{d y n, 3}^{i n}-\delta_3\right]}
	\end{array}\right).
\end{align}
By calculating the areas shown in Fig.~\ref{fig:Monkey-saddle-dynamical-phases} we find the following relations:
\begin{align}
	F_4=\text { const },\quad 
	F_3=\frac{2}{3} F_1,\quad
	F_2+2 F_3=F_4.
\end{align}
Next, we change the notation to $\phi_0=F_1 / 3$, and obtain
\begin{align}
	\left\{\begin{array}{l}
		\phi_{d y n, 3}^{i n}=-3 \phi_0 \\
		\phi_{d y n, 1}^{i n}=-F_4-2 \phi_0 \\
		\phi_{d y n, 2}^{i n}=F_4-\phi_0
	\end{array}, \quad\left\{\begin{array}{l}
		\phi_{d y n, 3}^{ {out }}=3 \phi_0 \\
		\phi_{d y n, 1}^{o u t}=-F_4+\phi_0 \\
		\phi_{d y n, 2}^{o u t}=F_4+2 \phi_0
	\end{array}\right.\right..
\end{align}
At the same time following the numerical procedure described in above section, we find S-matrix:
\begin{align}
	S^{\text{num}}=\left(\begin{array}{ccc}
		\mathcal{T}e^{i\varphi_t} e^{i 6 \phi_0(E)} &\mathcal{R}_2e^{i\varphi_2} e^{-i \frac{\pi}{3}+i 2 \phi_0(E)} &\mathcal{R}_1e^{i\varphi_1} e^{i \frac{\pi}{3}+i 4 \phi_0(E)} \\
		\mathcal{R}_1e^{i\varphi_1} e^{i \frac{\pi}{3}+i 4 \phi_0(E)} & \mathcal{T}e^{i\varphi_t} e^{i 6 \phi_0(E)} &\mathcal{R}_2e^{i\varphi_2} e^{-i \frac{\pi}{3}+i 2 \phi_0(E)} \\
		\mathcal{R}_2e^{i\varphi_2} e^{-i \frac{\pi}{3}+i 2 \phi_0(E)} & \mathcal{R}_1e^{i\varphi_1} e^{i \frac{\pi}{3}+i 4 \phi_0(E)} & \mathcal{T}e^{i\varphi_t} e^{i 6 \phi_0(E)}
	\end{array}\right).
\end{align}
Here, $\mathcal{T},\mathcal{R}_1,\mathcal{R}_2$ are real values. The presence of $e^{i\frac{\pi}{3}}$ and $e^{iF_4}$ phase factors is a consequence uncertainty of the dynamical phases, that are defined only up to an offset parameter. This arbitrariness is controlled by $e^{i\delta_i}$ phase factors which transform the numerical basis to the basis of the network model.  Phases $e^{\varphi_t},e^{\varphi_1}$ and $e^{\varphi_2}$ are scattering phases that decay to zero with $E \rightarrow \infty$. Therefore, we have
\begin{align}\label{eq:S-matrix-with-left-F}
	S=\left(\begin{array}{ccc}
		\mathcal{T}e^{i\varphi_t} e^{i 3 \phi_0(E)} & \mathcal{R}_2e^{i\varphi_2} e^{i\left(\delta_1-\delta_2+2 F_4-\frac{\pi}{3}\right)} & \mathcal{R}_1e^{i\varphi_1} e^{i\left(\delta_1-\delta_3+F_4+\frac{\pi}{3}\right)} \\
		\mathcal{R}_1e^{i\varphi_1} e^{i\left(\delta_2-\delta_1-2 F_4+\frac{\pi}{3}\right)} & \mathcal{T}e^{i\varphi_t} e^{i 3 \phi_0(E)} & \mathcal{R}_2e^{i\varphi_2} e^{i\left(\delta_2-\delta_3-F_4-\frac{\pi}{3}\right)-i 3 \phi_0(E)} \\
		\mathcal{R}_2e^{i\varphi_2} e^{i\left(\delta_3-\delta_1-F_4-\frac{\pi}{3}\right)-i 3 \phi_0(E)} & \mathcal{R}_1e^{i\varphi_1} e^{i\left(\delta_3-\delta_2+F_4+\frac{\pi}{3}\right)} & \mathcal{T}e^{i\varphi_t}
	\end{array}\right).
\end{align}
To determine the connection between $\delta_i$ and $F_4$, we may use the limiting cases of big energies where the $S$-matrix written in the network basis must take the form of full reflection scattering matrix for each of the three trajectories with additional factor $-i$ appearing at turning point \cite{Alexandradinata2018} (see Fig.~\ref{fig:triangular-lattice-orbit-notation}):
\begin{align}
	S(E \gg 0)=\left(\begin{array}{ccc}
		0 & 0 & -i \\
		1 & 0 & 0 \\
		0 & -i & 0
	\end{array}\right).
\end{align}
From this condition and previously obtained expression \eqref{eq:S-matrix-with-left-F}, we find a set of conditions, which in turn yield
\begin{align}
	\left\{\begin{array}{l}
		e^{i\left(\delta_2-\delta_1-2 F_4+\pi / 3\right)}=1 \\
		e^{i\left(\delta_3-\delta_2+F_4+\pi / 3\right)}=-i
	\end{array} \Rightarrow e^{i\left(\delta_1-\delta_3+F_4+\pi / 3\right)}=e^{i \pi} e^{-i\left(\delta_2-\delta_1-2 F_4+\pi / 3\right)} e^{-i\left(\delta_3-\delta_2+F_4+\pi / 3\right)}=-i\right. .
\end{align}
In such a way we settle all additional constants and obtained the exact form of the network model $S$-matrix from numerical $S^{\text{num}}$-matrix:
\begin{align}
	S = \begin{pmatrix}
		\mathcal{T}e^{i\varphi_t}e^{i3\phi_0(E)} & \mathcal{R}_2e^{i\varphi_2} & -i\mathcal{R}_1e^{i\varphi_1}\\
		\mathcal{R}_1e^{i\varphi_1} & \mathcal{T}e^{i\varphi_t}e^{i3\phi_0(E)} & i\mathcal{R}_2e^{i\varphi_2}e^{-i3\phi_0(E)}\\
		\mathcal{R}_2e^{i\varphi_2}e^{-i3\phi_0(E)} & -i\mathcal{R}_1e^{i\varphi_1} & \mathcal{T}e^{i\varphi_t}
	\end{pmatrix}.
\end{align}
Substituting it to \eqref{eq:scat} that couples solutions on neighboring orbits, we find:
\begin{align}
	\begin{pmatrix}
		\alpha_{1}\\
		\alpha_{2}\\
		\alpha_{3}
	\end{pmatrix} &=\begin{pmatrix}
		e^{i\left(p_k-\Phi_1\right)}&0&0\\
		0&e^{i\left(p_l+l_B^2\frac{b_{2,x}b_{2,y}}{2}\right)}&0\\
		0&0&1
	\end{pmatrix}\begin{pmatrix}
		\mathcal{T}e^{i\varphi_t}e^{i3\phi_0(E)} & \mathcal{R}_2e^{i\varphi_2} & -i\mathcal{R}_1e^{i\varphi_1}\\
		\mathcal{R}_1e^{i\varphi_1} & \mathcal{T}e^{i\varphi_t}e^{i3\phi_0(E)} & i\mathcal{R}_2e^{i\varphi_2}e^{-i3\phi_0(E)}\\
		\mathcal{R}_2e^{i\varphi_2}e^{-i3\phi_0(E)} & -i\mathcal{R}_1e^{i\varphi_1} & \mathcal{T}e^{i\varphi_t}
	\end{pmatrix}\nonumber\\
	&\times
	\begin{pmatrix}
		e^{-i\left(p_l+l_B^2\frac{b_{2,x}b_{2,y}}{2}\right)}&0&0\\
		0&e^{-i\Phi_2}&0\\
		0&0&e^{i\left(\Phi_3-p_k\right)}
	\end{pmatrix}\begin{pmatrix}
		\alpha_{1}\\
		\alpha_{2}\\
		\alpha_{3}
	\end{pmatrix}.
\end{align}
The condition for the existence of nontrivial solution, also known as non-linear eigenvalue problem of the Ho-Chalker operator \cite{HoChalker1996PRB}, gives the spectral equation
\begin{align}
	&\det S e^{-i(\Phi_1+\Phi_2-\Phi_3)/2}+e^{i(\Phi_1+\Phi_2-\Phi_3)/2}
	-((\mathcal{T}e^{i\varphi_t})^2e^{i6\phi_0}-\mathcal{R}_1e^{i\varphi_1}\mathcal{R}_2e^{i\varphi_2})\nonumber\\
	&\times\left[e^{-ip_l-i3\phi_0-i(\Phi_1-\Phi_2-\Phi_3+l_B^2b_{2,x}b_{2,y})/2}+e^{-i(p_k-p_l)-i3\phi_0-i(\Phi_2-\Phi_1-\Phi_3-\l_B^2b_{2,x}b_{2,y})/2}+e^{ip_k-i(\Phi_2+\Phi_1+\Phi_3)/2}\right]\nonumber\\
	&-\mathcal{T}e^{i\varphi_t} \left[e^{ip_l+i3\phi_0+i(\Phi_1-\Phi_2-\Phi_3+l_B^2b_{2,x}b_{2,y})/2}+e^{i(p_k-p_l)+i3\phi_0+i(\Phi_2-\Phi_1-\Phi_3-l_B^2b_{2,x}b_{2,y})/2}+e^{-ip_k+i(\Phi_2+\Phi_1+\Phi_3)/2}\right]=0.
\end{align}
Note that for this kind of problems the spectral equations are a convenient tool to solve eigenvalue problem of Ho-Chalker operator as it reduces to the well-known Lifshitz-Onsager quantization condition in crystals with small corrections coming from nonzero tunneling probabilities.  

\begin{figure*}
	\centering
	\includegraphics[scale=0.43]{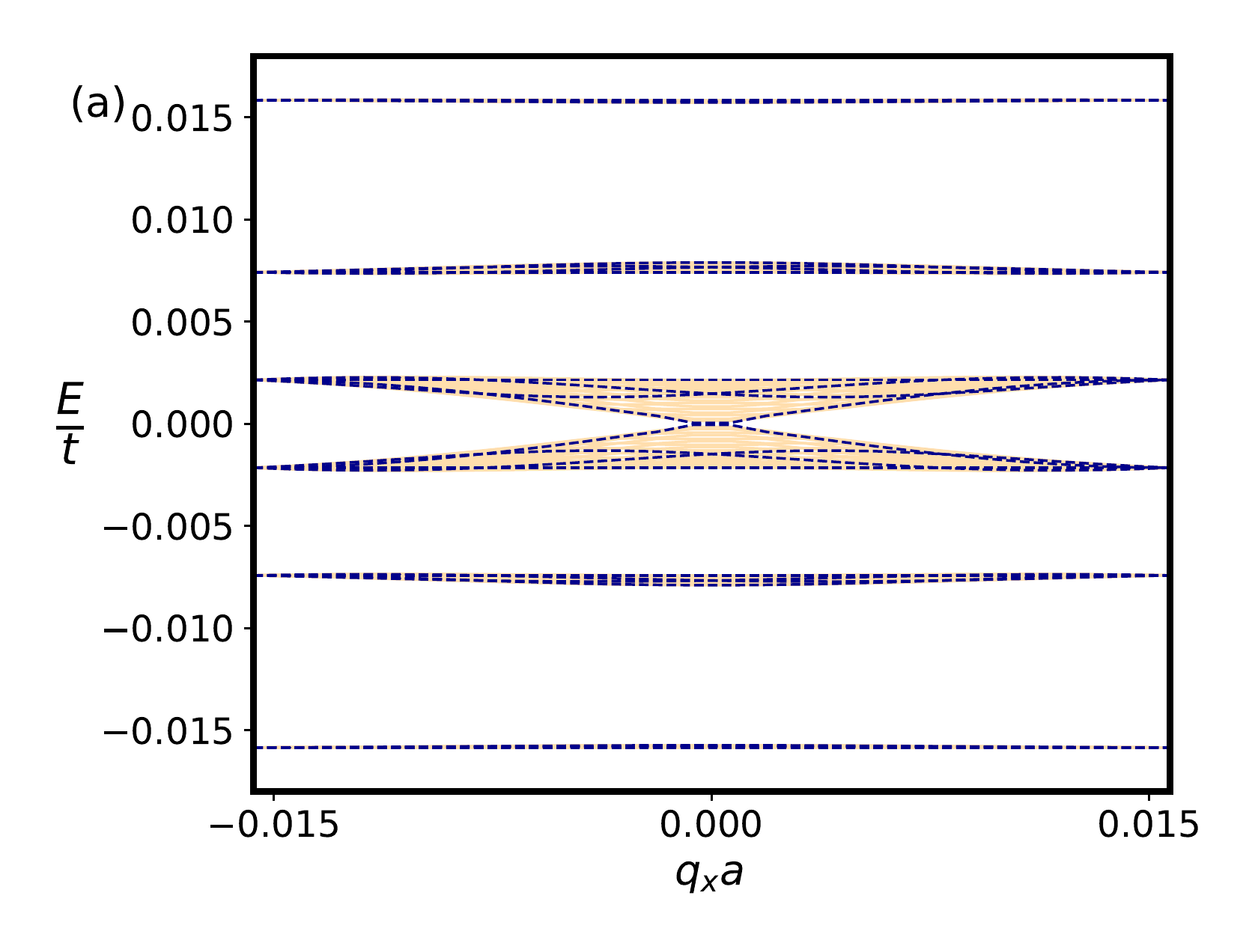}\qquad
	\includegraphics[scale=0.43]{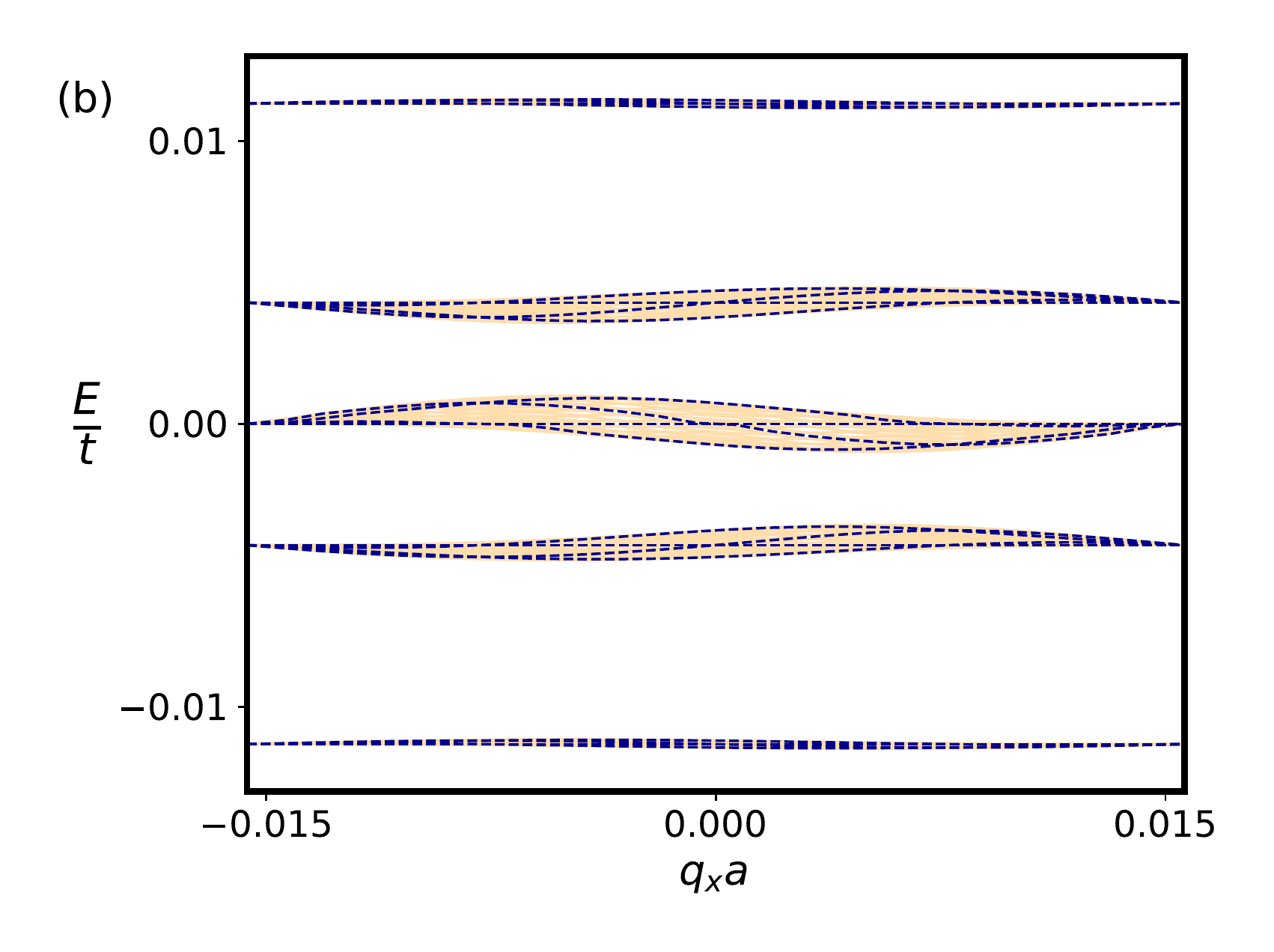}
	\caption{Two panels demonstrate comparison of analytic predictions (blue dashed lines) and numerical tight-binding simulations (orange solid lines merging into shaded areas) for the flux $\Phi=\frac{1}{p}\frac{h}{e}$ with (a) even $p=200$ and (b) odd $p=201$. The unit cell width of the ribbon with periodic boundary condition is taken to be $W=20\sqrt{3}p a$. }
	\label{fig:spectrum-triangular-supplement}
\end{figure*}

We note that from unitarity of $S$-matrix,
\begin{align}
	\mathcal{T}=|(\mathcal{T}e^{i\varphi_t})^2e^{i6\phi_0}-\mathcal{R}_1e^{i\varphi_1}\mathcal{R}_2e^{i\varphi_2}|, \quad \arg((\mathcal{T}e^{i\varphi_t})^2e^{i6\phi_0}-\mathcal{R}_1e^{i\varphi_1}\mathcal{R}_2e^{i\varphi_2})=\varphi_{sc}-\varphi_{t},
\end{align} where $\det S = e^{i\varphi_{sc}}$.
Therefore, we have
\begin{align}
	&e^{i\varphi_{sc}/2-i(\Phi_1+\Phi_2-\Phi_3)/2}+e^{i(\Phi_1+\Phi_2-\Phi_3)/2-i\varphi_{sc}/2}\nonumber\\
	&+ \mathcal{T}e^{i(\varphi_{sc}/2-\varphi_t)}\left[e^{-ip_l-i3\varphi_0-i(\Phi_1-\Phi_2-\Phi_3+l_B^2b_{2,x}b_{2,y})/2}+e^{-i(p_k-p_l)-i3\varphi_0-i (\Phi_2-\Phi_1-\Phi_3-l_B^2b_{2,x}b_{2,y})/2}+e^{ip_k-i(\Phi_2+\Phi_1+\Phi_3)/2}\right]\nonumber\\
	&+ \mathcal{T}e^{i(\varphi_t-\varphi_{sc}/2)} \left[e^{ip_l+i3\varphi_0+i(\Phi_1-\Phi_2-\Phi_3+l_B^2b_{2,x}b_{2,y})/2}+e^{i(p_k-p_l)+i3\varphi_0+i(\Phi_2-\Phi_1-\Phi_3-l_B^2b_{2,x}b_{2,y})/2}+e^{-ip_k+i(\Phi_2+\Phi_1+\Phi_3)/2}\right]=0.
\end{align}
Numerical calculations show that for the particular case of Monkey saddle point $\varphi_{sc} = 2\varphi_t$. Using geometry of the problem (see Fig.~\ref{fig:Monkey-saddle-dynamical-phases}), we note that the particular combinations of phases can be expressed via area enclosed by the orbit, $\Phi_1+\Phi_2-\Phi_3 = l_B^2\mathcal{A}(E) = \pi p - 2\Phi_3$ and $\Phi_3=2F_1=6\phi_0$. In addition, we note that $l_B^2 b_{1,x}b_{2,y}=\pi p$. Thus, we obtain a final form of spectral equation:
\begin{align}
	&\cos(l_B^2\mathcal{A}(E)/2-\varphi_{sc}/2)=\nonumber\\
	&=\mathcal{T}\Big[\cos(l_B^2[q_1b_{2,y}-q_2b_{2,x}]-\frac{\pi p}{2})+\cos\left(l_B^2[q_1b_{2,y}+q_2(b_{1,x}-b_{2,x})]-\frac{\pi p}{2}\right)+\cos\left(l_B^2 q_2 b_{1,x}-\frac{\pi p}{2}\right)\Big].
\end{align}
These spectral equations appears in the main text. For the sake of completeness, we present a difference in spectral structure with even and odd flux denominator values $p$ in Fig.~\ref{fig:spectrum-triangular-supplement}. As seen from the plots, for even $p=200$ spectrum is symmetric, while for odd $p=201$ spectrum is antisymmetric with momentum. The analytic predictions and numerical tight binding simulations demonstrate excellent agreement. 

\end{document}